\documentclass[acmsmall]{acmart}

\AtBeginDocument{%
  }

\acmDOI{XXXXXXX.XXXXXXX}

\acmJournal{JACM}
\acmVolume{1}
\acmNumber{1}
\acmArticle{1}
\acmMonth{1}

\usepackage{enumitem}
\usepackage{multirow}
\usepackage{graphicx}
\begin{document}

\title{MMGRec: Multimodal Generative Recommendation with Transformer Model}

\author{Han Liu}
\email{hanliu.sdu@gmail.com}
\affiliation{%
  \institution{School of Computer Science and Technology, Shandong University}
  \streetaddress{School of Computer Science and Technology}
  \city{Qingdao 266237}
  \country{China}
}

\author{Yinwei Wei}
\email{weiyinwei@hotmail.com}
\affiliation{%
  \institution{School of Software, Shandong University}
  \city{Jinan 250101}
  \country{China}}

\author{Xuemeng Song}
\email{sxmustc@gmail.com}
\affiliation{%
  \institution{Department of Computer Science and Engineering, Southern University of Science and Technology}
  \city{Shenzhen 518055}
  \country{China}
}

\author{Weili Guan}
\email{honeyguan@gmail.com}
\affiliation{%
  \institution{School of Information Science and Technology, Harbin Institute of Technology (Shenzhen)}
  \city{Shenzhen 518055}
  \country{China}}

\author{Yuan-Fang Li}
\email{yuanfang.li@monash.edu}
\affiliation{%
  \institution{Department of Data Science and AI, Monash University}
  \city{Clayton Victoria 3800}
  \country{Australia}}

\author{Liqiang Nie}
\authornote{Corresponding author.}
\email{nieliqiang@gmail.com}
\affiliation{%
  \institution{College of Informatics, Harbin Institute of Technology (Shenzhen)}
  \city{Shenzhen 518055}
  \country{China}}

\renewcommand{\shortauthors}{Han Liu et al.}

\begin{abstract}
Multimodal recommendation aims to recommend user-preferred candidates based on her/his historically interacted items and associated multimodal information. Previous studies commonly employ an \textbf{embed-and-retrieve} paradigm: learning user and item representations in the same embedding space, then retrieving similar candidate items for a user via embedding inner product. However, this paradigm suffers from inference cost, interaction modeling, and false-negative issues. Toward this end, we propose a new \textbf{MMGRec} model to introduce a \textbf{generative} paradigm into multimodal recommendation. Specifically, we first devise a hierarchical quantization method Graph RQ-VAE to assign Rec-ID for each item from its multimodal and CF information. Consisting of a tuple of semantically meaningful tokens, Rec-ID serves as the unique identifier of each item. Afterward, we train a Transformer-based recommender to generate the Rec-IDs of user-preferred items based on historical interaction sequences. The generative paradigm is qualified since this model systematically predicts the tuple of tokens identifying the recommended item in an autoregressive manner. Moreover, a relation-aware self-attention mechanism is devised for the Transformer to handle non-sequential interaction sequences, which explores the element pairwise relation to replace absolute positional encoding. Extensive experiments evaluate MMGRec's effectiveness compared with state-of-the-art methods.
\end{abstract}




\maketitle
\section{Introduction}
With the popularity of multimedia-centric scenarios like social media and micro-video websites, multimodal recommendation systems are gaining widespread attention and adoption. These systems generally follow the \textbf{embed-and-retrieve} paradigm. The paradigm first incorporates multimodal information with collaborative filtering (CF)~\cite{sarwar2001item} to learn representations for users and items in the embedding stage, and then locates the preferred items for users by measuring inner product similarity in the retrieval stage. Relevant research primarily focuses on enhancing representation learning in the embedding stage~\cite{liu2022disentangled,de2023disentangling}. For example, ACF~\cite{chen2017attentive} introduces a hierarchical attention mechanism to select informative modality content, while MMGCN~\cite{wei2019mmgcn} incorporates Graph Convolution Network (GCN) to achieve modal information aggregation and propagation.

Despite the remarkable performance, existing research overlooks the inherent issues caused by similarity calculation in the retrieval stage. First, similarity calculation becomes overwhelming. With $I$ items and $U$ users, capturing the top similar $K$ items for each user incurs time complexity $\mathcal{O}(UID+UI\text{log}K)$, where $D$ is the dimension of user/item representations. As users/items increase, the time consumption significantly grows, affecting recommendation efficiency. Secondly, linear inner product inadequately models user-item interactions' intricate structure~\cite{he2017neural}. Although some studies model non-linear interaction with neural networks~\cite{he2017neural} and metric learning~\cite{hsieh2017collaborative}, they sacrifice similarity calculation speed, exacerbating efficiency issues. Lastly, the paradigm assumes that interacted items are inherently closer to user preference than uninteracted ones~\cite{rendle2009bpr}, leading to a false-negative issue~\cite{wang2020reinforced} since the absence of interaction does not necessarily imply dislike~\cite{yang2023generate}. 

To address these issues, recent work TIGER~\cite{rajput2023recommender} introduces semantic IDs by inheriting the Differentiable Search Index algorithm~\cite{tay2022transformer} and proposes a \textbf{generative} paradigm-based recommendation framework. This framework aims to directly generate a list of user-preferred items without relying on similarity computation. To achieve this goal, two components, including \textbf{Semantic ID Assignment} and \textbf{Semantic ID Generation}, are designed to answer the questions: \textit{how to represent an item with a semantic ID} and \textit{how to predict the semantic IDs of preferred items for a user}. Initially, the method assigns each item a unique semantic ID --- a sequence of semantic tokens learned from its textual description. Then, a Transformer~\cite{vaswani2017attention} model is trained as the recommendation agent to predict the semantic ID of the next item based on the user's chronological interaction sequence. Each ID token is yielded with autoregressive decoding, and multiple item IDs are generated with beam search~\cite{sutskever2014sequence}. However, for multimodal recommendation, there are still two technical challenges untouched in the two components:
\begin{itemize}[leftmargin=*]
\item \textbf{Semantic ID Assignment:} Existing ID assignment approaches employ unimodal semantic clustering, inevitably leading to ID collision among similar items. Although random distinction~\cite{rajput2023recommender,tay2022transformer} is introduced to address this, such randomness information in ID is difficult to learn with Transformer, negatively affecting the performance of generative recommendation.
\item \textbf{Semantic ID Generation:} The position and order information in historical interaction sequences is important for Transformer to understand user behavior and preference. However, this information often lacks explicit references (\textit{e.g.}, timestamp, rating) and varies across different users. Therefore, it is challenging to make the Transformer aware of each item's position within the historical interaction sequence.
\end{itemize}

To overcome the challenges, we design a new ID structure --- \textbf{Rec-ID}, and present a Multimodal Generative Recommendation (\textbf{MMGRec}) framework for Rec-ID assignment and generation. Specifically, Rec-ID comprises a sequence of semantic tokens followed by a popularity token, leveraging the correlation between item popularity and semantic information. 
For Rec-ID assignment, building upon the RQ-VAE adopted in TIGER, we propose a Graph Residual-Quantized Variational AutoEncoder (Graph RQ-VAE) that incorporates a graph network to fuse item multimodal information with collaborative filtering signals, and quantizes the resulting representations into a sequence of semantic tokens.
Then, we rank the items with identical semantic tokens by popularity and use their ranking index as the last token to avoid the collision problem. After that, for Rec-ID generation, we train a Transformer-based model to directly generate the preferred items' Rec-IDs based on users' historical interactions. To capture item position information in historical interaction sequences, we design a relation-aware self-attention mechanism for Transformer.
By extending the original self-attention mechanism, this approach models user-specific relationships among items through personalized query, key, and value projection matrix parameters.
We conduct extensive experiments on three public datasets to demonstrate the rationality and effectiveness of the proposed MMGRec.

Overall, the main contributions of our work are three-fold:
\begin{itemize}[leftmargin=*]
\item We propose MMGRec, a novel Transformer-based recommendation framework that consists of Rec-ID assignment and generation. This is the first effort known to us to introduce the generative paradigm into multimodal recommendation.
\item Technically, we propose a multimodal information quantization framework, namely Graph RQ-VAE, which incorporates a popularity token for Rec-ID assignment. We further design a relation-aware self-attention mechanism within the Transformer architecture for Rec-ID generation.
\item We conduct empirical studies on three real-world datasets. Extensive results demonstrate that MMGRec achieves state-of-the-art performance with promising inference efficiency. The codes are released as a side contribution\footnote{https://github.com/hanliu95/MMGRec.}.
\end{itemize}
\section{Related Work}
\subsection{Multimodal Recommendation}
Research on multimodal recommendation aims to enhance recommender systems by leveraging multimodal information~\cite{SGFD2023MM}. Existing methodologies typically employ a combination of CF and content-based recommendation techniques~\cite{liu2023dynamic}. Early studies aim to integrate multimodal content for more rational representation learning. For example, the pioneering work VBPR~\cite{he2016vbpr} maps visual information to the representation space and concatenates it with CF representations. 
Subsequent work like ACF~\cite{chen2017attentive} and UVCAN~\cite{liu2019user} adaptively select multimodal features using different levels of attention mechanisms to learn representations of users and items. In recent years, the research trend turns to exploring the relationship between modal information and users/items. For instance, the representative work MMGCN~\cite{wei2019mmgcn} introduces a modality-aware GCN into multimodal recommendation tasks to aggregate and propagate multimodal information on the user-item graph. HHFAN~\cite{cai2021heterogeneous} constructs a modality-aware heterogeneous graph to model intricate interactions among users, micro-videos, and multimodal signals, thereby improving recommendation performance.

Despite their effectiveness, existing methods suffer from inherent limitations, such as overwhelming inference cost, insufficient interaction modeling, and false-negative issues, stemming from their embed-and-retrieve paradigm. To address these limitations, this study proposes a generative multimodal recommendation method.

\subsection{Transformer-Based Recommendation Model}
Since its inception, Transformer has demonstrated outstanding performance across various domains, such as natural language processing~\cite{wolf2020transformers} and computer vision~\cite{han2022survey}, yielding remarkable results. Owing to its proficiency in modeling contextual information, Transformer has emerged as a central focus in recommendation research. Researchers have developed Transformer-based recommendation models to capture item relationships and user behavior patterns in historical interactions~\cite{wu2020sse,xia2020multiplex}. PEAR~\cite{li2022pear} leverages the Transformer to capture interaction and contextual information from the history of clicked items. LightGT~\cite{wei2023lightgt} lightens the Transformer and combines it with GCN to enhance user preference modeling by considering item multimodal correlations. Recently, Transformer has emerged as a significant contributor to the development of generative recommendation. DSI~\cite{tay2022transformer} signifies a pivotal transition in information retrieval towards a generative paradigm, being the first successful end-to-end Transformer application for retrieval tasks. Relying on a novel semantic ID representation, TIGER~\cite{rajput2023recommender} implements a Transformer-based sequence-to-sequence model capable of directly predicting the semantic ID of the next item to achieve generative recommendation. 

Existing Transformer-based approaches are primarily limited to traditional representation learning or sequential generative recommendation. This study firstly introduces the Transformer with a generative paradigm to the field of multimodal recommendation.

\subsection{LLM-Based Recommendation Systems}
In recent years, recommendation systems have increasingly incorporated Large Language Models (LLMs) to exploit their rich knowledge and strong reasoning abilities, mainly in two paradigms: LLM-augmented and LLM-based recommendation systems~\cite{wu2024survey}.

LLM-augmented recommendation systems typically use LLM-generated embeddings as semantic representations~\cite{hou2023learning} or high-level feature extractors~\cite{xi2024towards}, while still relying on traditional models for final decision making, which limits the use of LLMs’ reasoning ability~\cite{hou2022towards}. In contrast, LLM-based recommenders directly perform recommendation with LLMs. Early work exploited in-context learning for zero- or few-shot recommendation~\cite{dai2023uncovering}, but untuned models suffer from weak instruction following and limited domain knowledge. Recent studies, therefore, adopt supervised fine-tuning with historical interaction data to improve recommendation performance~\cite{bao2023tallrec,lin2024data}.

Our approach is conceptually closer to the second line of work, but differs in that we adopt a from-scratch training paradigm. Compared with fine-tuning large pre-trained models, fully training a small generative recommender offers better efficiency, stronger domain adaptation, and clearer controllability, making it more suitable for scalable and resource-constrained recommendation scenarios.

\section{Preliminary}

This work is inspired by TIGER (short for \textit{Transformer Index for GEnerative Recommenders}), a recent generative model for sequential recommendation~\cite{rajput2023recommender}. Its foundation lies in a novel semantic ID for item identification, consisting of an ordered semantic token tuple:
\begin{equation}
    (c_{1}, \cdots, c_{M-1})=\phi(\mathbf{f}^t),
\end{equation}
where $\mathbf{f}^t$ is the item's textual feature converted from descriptions using Sentence-T5~\cite{ni2022sentence}. $\phi(\cdot)$ indicates the vector quantization algorithm~\cite{zeghidour2021soundstream} converting $\mathbf{f}^t$ into a tuple of $M-1$ semantic tokens.

To address the collision problem that items with similar textual features tend to be assigned the same tokens,  the method inserts an extra token at the end of the semantic tokens to ensure the uniqueness of IDs. Specifically, given $K$ items with the same tokens $(c_{1}, \cdots, c_{M-1})$, it randomly sorts the items and sets their corresponding indices in the sorted order as the extra tokens:
\begin{equation}
    (p_1, p_2, \cdots, p_K)= \text{argsort}_{rand}(\text{item}_1, \text{item}_2, \cdots, \text{item}_K),
\end{equation}
where $\text{argsort}_{rand}(\cdot)$ is the random ranking function to sort $K$ items. $p_{k}\in\{1, 2, \cdots, K\}$ denotes the index of the $k$-th item in the order. Then, after concatenating them at the end of semantic tokens, the semantic ID of item $k$ can be obtained as:
\begin{equation}
    S_k = (c_{1}, \cdots, c_{M-1}, p_k).
\end{equation}

With the obtained semantic ID, a Transformer-based sequence-to-sequence model is trained on sequences of semantic IDs derived from a user’s chronological item interactions.
The objective is to predict the semantic ID of the next item after the sequence. This model autoregressively decodes the tokens of semantic ID identifying the next item,  thereby qualifying as a generative recommendation. 

However, this method has notable limitations. The random token lacks semantic meaning and is challenging to model using statistical machine learning techniques like Transformer. Additionally, the method struggles with the input of non-sequential interactions.
\section{Method}

In this section, we first elaborate on our designed MMGRec model, which consists of two stages: Rec-ID assignment and Rec-ID generation. After that, we detail the model training and inference, followed by a discussion regarding the time complexity.

\subsection{Rec-ID Definition}  
As the foundation of generative recommendation, we define a new item identifier Rec-ID equipped with two attributes:
1) \textbf{Semantics}, where the Rec-ID condenses the semantic information of the item, and 2) \textbf{Uniqueness}, where the Rec-ID is capable of distinguishing each item from others. 

Towards this end, Rec-ID is designed to consist of a sequence of semantic tokens and an extra token based on the item's popularity. The motivation is that the popularity of an item is relevant to its semantic information. Hence, during the autoregressive generation phase, the last token (\textit{i.e.,} popularity token) can be predicted by jointly analyzing the semantic tokens generated before.

\subsection{Rec-ID Assignment} 
According to the design of Rec-ID, we detail the Rec-ID assignment to describe how to allocate the semantic tokens and the popularity token for each item.
\subsubsection{\textbf{Graph RQ-VAE}}
\begin{figure}[tbp]
\centering
\includegraphics[width=0.55\linewidth]{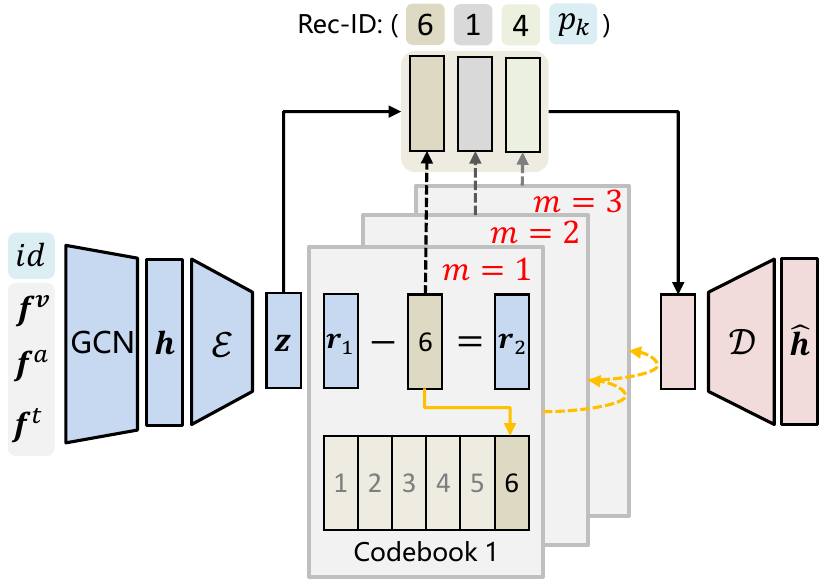}
\caption{An illustration of Graph RQ-VAE model architecture.}
\label{fig:rqvae}
\end{figure}

To learn the semantic tokens for items, as shown in Figure~\ref{fig:rqvae}, we devise a Graph RQ-VAE model to fuse their multimodal information and further quantize into codewords. Alongside multimodal (\textit{e.g.}, visual, acoustic, and textual) features, historical user-item interactions are fed to this model. These interactions not only aid in extracting relevant multimodal features for recommendation~\cite{wei2023lightgt} but also reveal behavioral signals related to item popularity. In particular, a user-item bipartite graph is constructed, where nodes represent users and items, and edges denote interactions. 
To fuse these heterogeneous inputs into a unified representation, we initialize item $i$ as follows:
\begin{equation}    \mathbf{h}_i^{(0)}=\mathbf{h}_i^m\oplus\mathbf{h}_i^c, \quad \mathbf{h}_i^{m}=(\mathbf{f}_i^v\oplus\mathbf{f}_i^a\oplus\mathbf{f}_i^t)\mathbf{W},
\end{equation}
where $\oplus$ denotes the vector concatenation operation.
$\mathbf{h}_i^{m}\in\mathbb{R}^{D}$ represents item $i$'s multimodal embedding learned by concatenating its visual, acoustic, and textual features, \textit{i.e.,} $\mathbf{f}_i^v\in\mathbb{R}^{D^v}$, $\mathbf{f}_i^a\in\mathbb{R}^{D^a}$, and $\mathbf{f}_i^t\in\mathbb{R}^{D^t}$. $D^v$, $D^a$, and $D^t$ denote the dimensions of these modality features. $\mathbf{W}\in\mathbb{R}^{(D^v+D^a+D^t) \times D}$ is a parameter matrix that maps multimodal features to the $D$-dimensional embedding space. Combining this multimodal embedding with a randomly initialized CF embedding $\mathbf{h}_i^c\in\mathbb{R}^{D}$ yields the item representation $\mathbf{h}_i^{(0)}\in\mathbb{R}^{2D}$. For performing graph convolution operations on the bipartite graph, user representations are randomly initialized~\cite{lei2023learning}, such as $\mathbf{h}_u^{(0)}\in\mathbb{R}^{2D}$ for user $u$.  

Treating them as node embeddings at the $0$-th layer, item $i$’s representation at the $y$-th GCN layer is obtained as:
\begin{equation}
    \mathbf{h}_i^{(y)}=\text{LeakyReLU}(\mathbf{h}_i^{(y-1)}\mathbf{W}_1 + \frac{1}{|\mathcal{N}_i|}\sum_{u\in\mathcal{N}_i}\mathbf{h}_u^{(y-1)}\mathbf{W}_2),
\end{equation}
where $\mathcal{N}_i$ is the neighborhood set of item $i$ (\textit{i.e.}, users whom item $i$ directly interacted with), and LeakyReLU is the activation function. Analogously, the representation $\mathbf{h}_u^{(y)}$ for user $u$ is obtained by information propagation from its neighbor items. After stacked GCN layers, final representations are denoted as $\mathbf{h}_u\in\mathbb{R}^{D}$ and $\mathbf{h}_i\in\mathbb{R}^{D}$. Note that we utilize the concise graph convolution operation~\cite{hamilton2017inductive}, leaving complicated choices~\cite{liu2023dynamic} for further exploration.

Based on the final item representations, we employ Residual-Quantized Variational AutoEncoder (RQ-VAE)~\cite{zeghidour2021soundstream} to generate semantic tokens of Rec-IDs. As a multi-stage vector quantizer, RQ-VAE can produce a tuple of codewords by quantizing residuals across hierarchical levels. As illustrated in Figure~\ref{fig:rqvae}, we input the learned item representation into an encoder model:  
\begin{equation}
    \mathbf{z}_i = \mathcal{E}(\mathbf{h}_i),
\end{equation}
where $\mathbf{z}_i$ is the output latent vector and $\mathcal{E}$ is the encoder, comprising a two-layer Multi-Layer Perceptron (MLP). At the first level, the initial residual is $\mathbf{r}_{i,1}=\mathbf{z}_i$. A codebook $\mathbf{B}_1=\{\mathbf{b}_l\}_{l=1}^L$ is employed, where $\mathbf{b}_l$ denotes the parameter embedding and $L$ represents the codebook size. Then, $\mathbf{r}_{i,1}$ is quantized by mapping it to the closest embedding from this codebook:
\begin{equation}
    c_{i,1}=\text{argmin}_l||\mathbf{r}_{i,1}-\mathbf{b}_l||_2^2,
\end{equation}
where $c_{i,1}$ is the index of the closest embedding, \textit{i.e.}, the first codeword. Recursively, for subsequent level $m$, the residual is defined as $\mathbf{r}_{i,m}=\mathbf{r}_{i,m-1}-\mathbf{b}_{c_{i,m-1}}$, and the codeword $ c_{i, m}$ is similarly computed using the level-specific codebook $\mathbf{B}_m$. By iteratively performing $M-1$ operations, a sequence of codewords is generated as follows:
\begin{equation}
    (c_{i, 1}, \cdots, c_{i, M-1})=\text{RQ-VAE}(\mathbf{h}_i),
\end{equation}
where $ c_{i, m}$ denotes the $m$-th codeword used as the $m$-th semantic token in Rec-ID of item $i$. Notably, the recursive approach approximates the input in a coarse-to-fine manner. Since each codeword is sourced from a distinct codebook, the capacity of Rec-IDs to uniquely represent items equals the product of all codebook sizes.

Upon the codewords, we use a two-layer MLP decoder $\mathcal{D}$ to reconstruct the input $\mathbf{h}_i$ as follows:
\begin{equation}
    \widehat{\mathbf{h}}_i=\mathcal{D}(\mathbf{z}_i+\text{sg}(\widehat{\mathbf{z}}_i-\mathbf{z}_i)),\quad\widehat{\mathbf{z}}_i=\sum_{m=1}^{M-1}\mathbf{b}_{c_{i,m}},
\end{equation}
where $\widehat{\mathbf{h}}_i$ denotes the decoder output, and $\widehat{\mathbf{z}}_i$ is the quantized representation of $\mathbf{z}_i$. $\text{sg}(\cdot)$ represents the stop-gradient operation. The loss function for training Graph RQ-VAE is as follows:
\begin{equation}\label{eq:}
\begin{split}   &\mathcal{L}=\mathcal{L}_{\text{bpr}}+\mathcal{L}_{\text{rqvae}}, \\ \mathcal{L}_{\text{bpr}}=&\sum_{(u,i,j)\in\mathcal{R}} -\text{ln}~\sigma(\mathbf{h}_u\mathbf{h}_i^\top-\mathbf{h}_u\mathbf{h}_j^\top),\\
    \mathcal{L}_{\text{rqvae}}=\sum_{i=1}^{I}\big(||\mathbf{h}_i-\widehat{\mathbf{h}}_i||^2 &+
    \sum_{m=1}^{M-1}(||\text{sg}(\mathbf{r}_{i,m})-\mathbf{b}_{c_{i,m}}||^2 + \beta||\mathbf{r}_{i,m}-\text{sg}(\mathbf{b}_{c_{i,m}})||^2)\big).
\end{split}
\end{equation}
Here $\mathcal{L}_{\text{bpr}}$ is the pairwise BPR~\cite{rendle2009bpr} loss to learn user and item representations. $\mathcal{R}=\{(u,i,j)|(u,i)\in\mathcal{R}^+,(u,j)\in\mathcal{R}^-\}$, where $\mathcal{R}^+$ indicates the observed interactions, and $\mathcal{R}^-$ indicates the unobserved interactions. $\sigma(\cdot)$ is the Sigmoid function. The loss function $\mathcal{L}_{\text{rqvae}}$ facilitates the simultaneous training of the encoder-decoder and codebooks in RQ-VAE. $I$ denotes the number of items. We alternatively optimize one of two sub-losses to learn model parameters.

\subsubsection{\textbf{Handling Collisions}}
To handle the collision problem, we expand a popularity token at the end of the semantic tokens to ensure Rec-IDs' uniqueness. In particular, given $K$ items with the same tokens $(c_{1}, \cdots, c_{M-1})$, we sort the items according to their popularity, \textit{i.e.}, the total number of interactions with users. The corresponding indices in the sorted order are used as their popularity tokens, formally, 
\begin{equation}
    (p_1, p_2, \cdots, p_K)= \text{argsort}_{pop}(\text{item}_1, \text{item}_2, \cdots, \text{item}_K),
\end{equation}
where $\text{argsort}_{pop}(\cdot)$ is the popularity ranking function to sort $K$ items. $p_{k}\in\{1, 2, \cdots, K\}$ denotes the index of the $k$-th item in the order. Then, after concatenating them at the end of semantic tokens, the Rec-ID of item $k$ can be obtained as: 
\begin{equation}
    R_k = (c_{1}, \cdots, c_{M-1}, p_k).
\end{equation}

\begin{figure}[!tbp]
\centering
\includegraphics[width=0.51\linewidth]{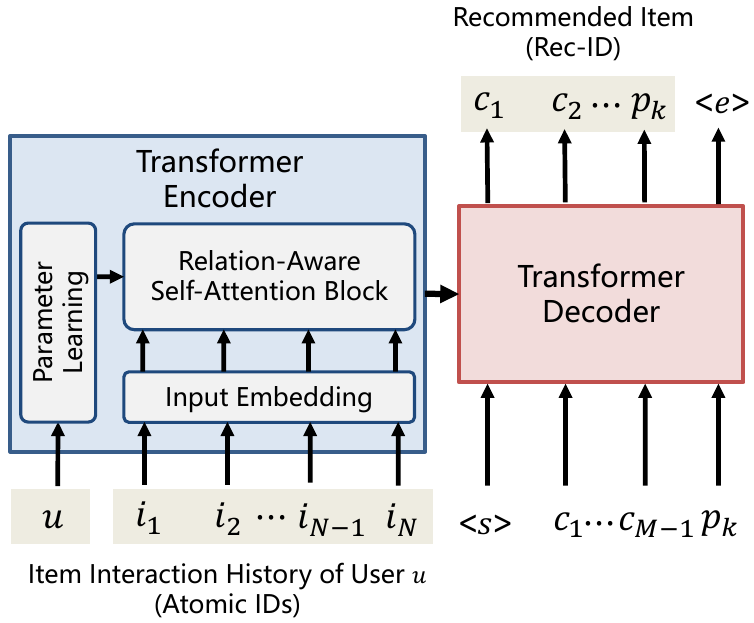}
\caption{Schematic illustration of Transformer encoder-decoder setup for building our generative recommendation model.}
\label{fig:framework}
\end{figure}

\subsection{Rec-ID Generation}
\subsubsection{\textbf{Transformer Input Embedding}}
To implement generative recommendation, a sequence-to-sequence Transformer model is exploited as a recommender. As shown in Figure~\ref{fig:framework}, the model consists of stacked encoder and decoder layers. To adopt the model in recommendation tasks, we organize historically interacted items of each user as a sequence of tokens and feed them into the encoder. 

Taking user $u$ as an example, we gather her/his interacted items' atomic IDs together as the Transformer's input sequence, \textit{e.g.}, $(i_1, \cdots, i_N)$. Then, we utilize the learned item representations for input embedding to convert the input tokens to vectors, formally, 
\begin{equation}
\begin{bmatrix}\mathbf{e}_{1},\cdots,\mathbf{e}_{N}\end{bmatrix}=\begin{bmatrix}\mathbf{h}_{i_1},\cdots,\mathbf{h}_{i_N}\end{bmatrix}=\text{Emb}(i_1, \cdots, i_N),
\end{equation}
where $\mathbf{e}_{i}\in\mathbb{R}^{D}$ denotes the embedding of the corresponding item in the sequence. 
It is worth noting that the input sequence length is uniformly set to $N$, filling the missing positions with padding token $0$. 

\subsubsection{\textbf{Relation-Aware Self-Attention}}
Following input embedding, positional encoding in Transformer is crucial for providing order information of the input sequence. Unlike sequential recommendation tasks, the input interacted item sequence is not well-ordered in our case, and its actual order implicitly and dynamically changes with different users. Thus, traditional absolute positional encoding is ineffective in this context. To overcome the problem, we develop a relation-aware self-attention mechanism for the Transformer. This mechanism explicitly models user-specific pairwise relations among input items based on user preferences, thereby encoding relative position information within interaction sequences.

\begin{figure}[tbp]
\centering
\includegraphics[width=0.7\linewidth]{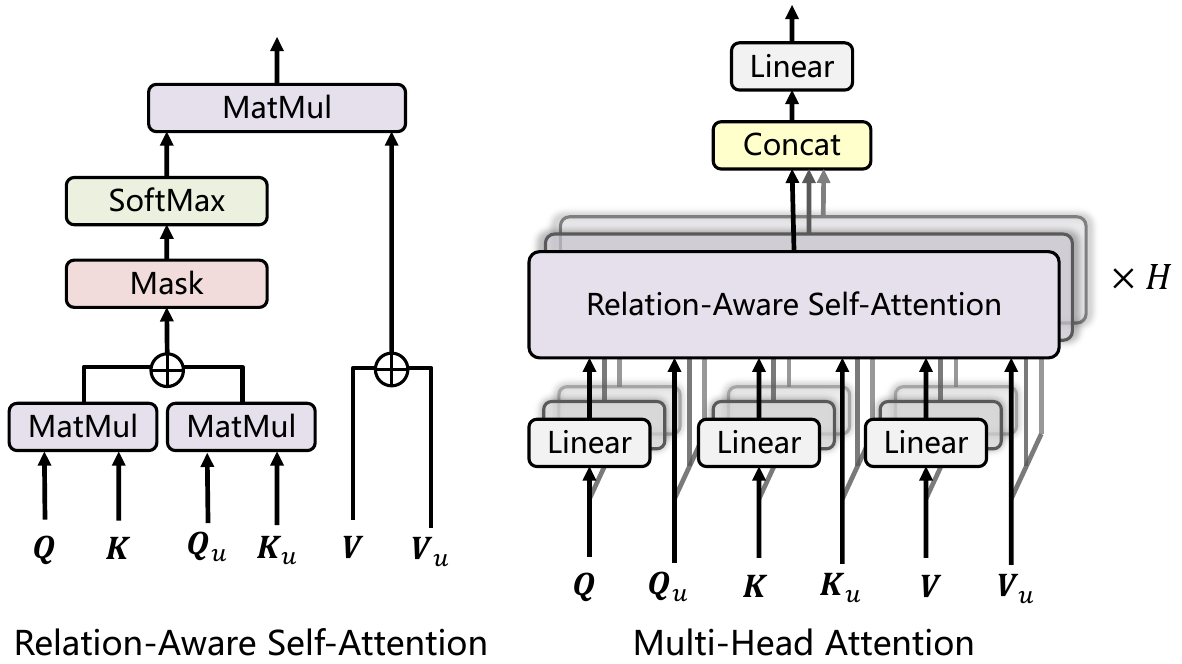}
\caption{An illustration of relation-aware self-attention and its corresponding multi-head attention. Therein, $\mathbf{Q}=\mathbf{E}\mathbf{W}^Q$, $\mathbf{Q}_u=\mathbf{E}\mathbf{W}^Q_u$, and the remaining parameters are computed analogically.}
\label{fig:attn}
\end{figure}

Specifically, as shown in Figure~\ref{fig:attn}, we extend the self-attention sublayer by incorporating user-specific parameters. First, we introduce the user-specific position encoder, which aims to encode the position information of each item from different users' perspectives:
\begin{equation}
\mathbf{p}_{j}^u=\mathbf{e}_j\mathbf{W}_u^V,
\end{equation}
where $\mathbf{W}_u^V$ represents the parameter matrix of user $u$'s position encoder and $\mathbf{p}_{j}^u$ is the encoded vector of item $j$. 

After obtaining the encoded vectors, we conduct the multi-head attention operations on the input sequence. In each head, we map the input element embeddings $\mathbf{E}=\begin{bmatrix}\mathbf{e}_{1},\cdots,\mathbf{e}_{N}\end{bmatrix}$ into query, key, and value spaces, and compute their new embeddings $\mathbf{X}=\begin{bmatrix}\mathbf{x}_{1},\cdots,\mathbf{x}_{N}\end{bmatrix}$ where $\mathbf{x}_{i}\in\mathbb{R}^{D}$: 
\begin{equation}\label{eq:new_z}
    \mathbf{x}_i = \sum_{j=1}^{N}\alpha_{ij}(\mathbf{e}_j\mathbf{W}^V + \mathbf{p}_{j}^u),
\end{equation}
where $\mathbf{W}^V$ is the matrix mapping the elements to the value space. $\alpha_{ij}$ is the weight coefficient computed with the Softmax function:
\begin{equation}
    \alpha_{ij} = \frac{\text{exp}(\epsilon_{ij})}{\sum_{n=1}^N\text{exp}(\epsilon_{in})},
\end{equation}
wherein $\epsilon_{ij}$ is calculated through a compatibility function that compares two input elements and considers their pairwise relation in the context of user $u$:
\begin{equation}\label{eq:new_a}
    \epsilon_{ij} = \frac{(\mathbf{e}_i\mathbf{W}^Q)(\mathbf{e}_j\mathbf{W}^K)^\top  + (\mathbf{e}_i\mathbf{W}_u^Q)(\mathbf{e}_j\mathbf{W}_u^K)^\top}{\sqrt{D}},
\end{equation}
where $\mathbf{W}^Q$ and $\mathbf{W}^K$ are parameter matrices mapping elements to the query and key spaces, respectively. Moreover, elements are also mapped into user-specific query and key spaces to compute the relation between each item pair. $\mathbf{W}_u^Q$ and $\mathbf{W}_u^K$ denote two user-specific parameter matrices for this space transition.



Considering the user number, adopting independent parameter matrices for each user will result in the issue of model parameter overload. Hence, we alternatively leverage user information to fine-tune a series of foundational parameters. Taking $\mathbf{W}_u^V$ as an example, its acquisition is as follows:

\begin{equation}
    \mathbf{W}_u^V = \text{MLP}(\mathbf{h}_u)\cdot\mathbf{U}^V,
\end{equation}
where $\mathbf{h}_u$ is the personal representation of user $u$ learned in Graph RQ-VAE. It is projected through MLP into a scalar, measuring the bias introduced by user $u$ on the foundational matrix parameter $\mathbf{U}^V\in\mathbb{R}^{D\times D}$. This strategy allows regulating the number of model parameters while maintaining user-specificity. $\mathbf{W}_u^Q$ and $\mathbf{W}_u^K$ can be obtained in the same way.

Notably, the relation-aware self-attention can be efficiently computed using the scaled dot product: 
\begin{equation}
    \mathbf{X} = \text{Softmax}(\frac{\mathbf{E}({\mathbf{W}^Q}{\mathbf{W}^K}^\top + {\mathbf{W}_u^Q}{\mathbf{W}_u^K}^\top)\mathbf{E}^\top}{\sqrt{D}})\mathbf{E}(\mathbf{W}^V+\mathbf{W}_u^V).
\end{equation}
The parameter matrices $\mathbf{W}^Q$, $\mathbf{W}^K$, $\mathbf{W}^V\in\mathbb{R}^{D\times D}$ are unique to each attention head, contributing sufficient expressive capacity. While the user-specific parameters $\mathbf{W}_u^Q$, $\mathbf{W}_u^K$, $\mathbf{W}_u^V\in\mathbb{R}^{D\times D}$ can be shared across all attention heads. 
To form the sublayer output, the outcomes from each head are concatenated, and a linear transformation is applied.

In our overall architecture, we employ a Transformer encoder with relation-aware self-attention. The decoder follows the standard Transformer design for autoregressive Rec-ID generation.

\subsection{Training \& Inference}

During training, each interaction ($u, i$) is treated as a training instance. The encoder takes the atomic ID-based interaction sequence of user $u$ as input, excluding the target item $i$. The decoder is trained to autoregressively generate the Rec-ID of item $i$ using a cross-entropy loss, where the Graph RQ-VAE codebook indices serve as the output vocabulary.
During inference, the trained Transformer generates Rec-IDs in an autoregressive manner. We apply beam search~\cite{sutskever2014sequence} to maintain the top-$K$ candidate sequences at each decoding step, yielding $K$ complete Rec-IDs after $M$ steps for top-$K$ recommendation. To avoid invalid Rec-ID generation, we further adopt constrained beam search~\cite{de2020autoregressive}, where the candidate token set at each step is restricted based on valid prefixes.

\subsection{Discussion}
Moreover, we discuss the inference time complexity of MMGRec compared with the most efficient method of traditional paradigm, \textit{i.e.}, Matrix Factorization (MF)~\cite{koren2009matrix}. Assuming the existence of $I$ items, each represented by $D$-dimensional embeddings, MF requires a time complexity of $\mathcal{O}(ID)$ to complete all inner product computations and $\mathcal{O}(I\text{log}K)$ to identify the top-$K$ recommendation for a single user. In contrast, MMGRec utilizes autoregressive decoding with KV caching and beam search for inference, where the beam width is set to $K$. For an $H$-layer MMGRec, generating $K$ sequences of length $M$ requires $\mathcal{O}(KH(M^2D+MD^2))$ time. Additionally, vocabulary projection and beam selection over $M$ decoding steps introduce complexities of $\mathcal{O}(MKDL)$ and $\mathcal{O}(MKL\text{log}K)$, respectively, where $L$ denotes the vocabulary size. Therefore, the overall inference complexity of MMGRec is $\mathcal{O}(KH(M^2D+MD^2)+MKDL+MKL\text{log}K)$. This analysis suggests that the generative paradigm may reduce inference time when the number of items is sufficiently large, while the Rec-ID length, beam width, and vocabulary size remain relatively small.


\section{Experimental Setup}

\subsection{Datasets \& Evaluation Metrics}
To evaluate the effectiveness of our proposed model, we conduct extensive experiments on three benchmark datasets: MovieLens\footnote{https://movielens.org/.}, TikTok\footnote{https://www.tiktok.com/.}, and Kwai\footnote{https://www.kwai.com/.}, which are widely used in multimodal recommendation study. The dataset statistics are summarized in Table~\ref{tab:dataset}.

\begin{itemize}[leftmargin=*]
\item \textbf{MovieLens}: This dataset was created for research by MMGCN authors. They collected movie trailers and descriptions based on the original dataset. 
Visual, acoustic, and textual features are extracted from frames, audio tracks, and descriptions using pre-trained models~\cite{he2016deep,hershey2017cnn,arora2017simple}, respectively.
\item \textbf{TikTok}: This dataset was released by the short-video sharing platform TikTok, including user-item interactions and multimodal features (visual, acoustic, textual) of items. According to the publisher's statement, the multimodal features are extracted from the raw data of short videos.
\item \textbf{Kwai}: The dataset was released by the short-video sharing platform Kwai, comprising user behavioral records and short videos' multimodal features (excluding acoustic features). Likewise, the multimodal features are obtained by pre-trained extractors.
\end{itemize}

\begin{table}[]
    \centering
    \caption{Statistics of the evaluation datasets. ($D^v$, $D^a$, and $D^t$ denote the dimensions of visual, acoustic, and textual modality feature data, respectively.)}
    \begin{tabular}{c|c|c|c|c|c|c}
         \hline
         Dataset & \#Users & \#Items & \#Inter. & $D^v$ & $D^a$ & $D^t$ \\
         \hline
         \hline
         MovieLens & 55,485 & 5,986 & 1,239,508 &  2048 & 128 & 100\\
         \hline
         TikTok & 36,656 & 76,085 & 726,065 &  128 & 128 & 128\\
         \hline
         Kwai & 7,010 & 86,483 & 298,492 &  2048 & - & 128\\
         \hline
    \end{tabular}
    \label{tab:dataset}
    \vspace{-0.45cm}
\end{table}

For each dataset, we randomly select 80\% of the historical interactions per user to constitute the training set, 10\% for validation, and the remaining 10\% for testing. The validation and testing sets are used for hyper-parameter tuning and performance evaluation, respectively. We adopt Recall (Recall@$K$) and Normalized Discounted Cumulative Gain (NDCG@$K$) as evaluation metrics for top-$K$ recommendation and preference ranking. Defaulting $K=10$, we report the average metric value for all users in the testing set.

\subsection{Baselines}
To demonstrate the effectiveness, we compare our method with the following state-of-the-art baselines, briefly divided into CF-based (\textit{i.e.}, GraphSAGE, NGCF, GAT, and LightGCN) and multimodal (\textit{i.e.}, VBPR, MMGCN, GRCN, LATTICE, InvRL, and LightGT).
\begin{itemize}[leftmargin=*]
\item \textbf{GraphSAGE}~\cite{hamilton2017inductive} passes information along the graph structure and aggregates them to update each node's representation.
\item \textbf{NGCF}~\cite{wang2019neural} encodes CF signals into representation learning by leveraging high-order connectivity from user-item interactions.
\item \textbf{GAT}~\cite{velivckovic2018graph} automatically learns weights for each node's neighbors and alleviates noisy information to improve GCN performance.
\item \textbf{LightGCN}~\cite{he2020lightgcn} simplifies GCN components, utilizing a weighted sum aggregator as the graph convolution operation.
\item \textbf{VBPR}~\cite{he2016vbpr} integrates multimodal information into the MF framework to predict user-item interactions.
\item \textbf{MMGCN}~\cite{wei2019mmgcn} assigns a dedicated GCN to each modality to learn modality-specific user preferences and item representations through modality-wise information propagation.
\item \textbf{GRCN}~\cite{wei2020graph} adaptively adjusts the interaction graph's structure according to the model training status, then applies graph convolution layers to distill informative signals on user preference.
\item \textbf{LATTICE}~\cite{zhang2021mining} discovers item relationships via multimodal features to construct a graph structure, on which graph convolutions aggregate high-order affinities for recommendation.
\item \textbf{InvRL}~\cite{du2022invariant} eliminates spurious correlations via heterogeneous environments to learn consistent invariant item representations across diverse settings, improving predictions of interactions.
\item \textbf{LightGT}~\cite{wei2023lightgt} is a state-of-the-art Transformer-based model. It designs modal-specific embedding and layer-wise position encoding for effective feature distillation, learning high-quality user preferences on item content for interaction prediction.
\end{itemize}

To ensure consistency, we adopt the publicly available implementations of the baselines.
Note that we do not include TIGER~\cite{rajput2023recommender} as a baseline. TIGER is designed for sequential recommendation and requires temporally ordered user-item interactions. In contrast, our work focuses on non-sequential collaborative filtering, and the datasets used in our experiments do not contain temporal labels. Moreover, TIGER constructs semantic IDs solely from textual features, while our method relies on multimodal item representations. Therefore, TIGER is not directly applicable to our setting.

\subsection{Parameter Settings}
We implement our MMGRec using Pytorch\footnote{https://pytorch.org/.} and Pytorch Geometric\footnote{https://pytorch-geometric.readthedocs.io/en/latest/.}. Model parameters are initialized with Xavier approach~\cite{glorot2010understanding} and optimized using the SGD optimizer~\cite{bottou2010large} with a batch size selected from \{500, 1,000, 2,000, 3,000\}. Hyper-parameters are tuned through grid search based on the results from the validation set. The optimal learning rate is searched from \{0.0001, 0.0005, 0.001, 0.005\} and ultimately set to 0.001. The $L_2$ normalization coefficient is searched within \{$10^{-6}$, $10^{-5}$, · · ·, $10^{-1}$, 1\}, with the optimal value set to $10^{-5}$. Moreover, early stopping is adopted if Recall@$10$ on the validation set does not rise for 20 successive epochs. Without specification, user/item representations default to a size of  $64$ in all methods for fairness. As for the Graph RQ-VAE model, we adopt three-stage quantization to assign a three-tuple Rec-ID for each item. To avoid Rec-ID collision, we add a popularity token as the fourth. The size of each level codebook is tuned in \{64, 128, 256\}. When computing the loss $\mathcal{L}_\text{rqvae}$, we set $\beta=0.25$.

\section{Experimental Results}
We conduct extensive quantitative and qualitative experiments to answer the following research questions:
\begin{itemize}[leftmargin=*]
\item \textbf{RQ1}: How does MMGRec perform on the multimodal recommendation task compared with state-of-the-art baselines?
\item \textbf{RQ2}: Is the technical choice of each component (\textit{i.e.}, item representations, Graph RQ-VAE, and relation-aware self-attention)  effective for MMGRec?
\item \textbf{RQ3}: How do hyper-parameter settings (\textit{e.g.}, depth of layer and number of heads) affect MMGRec?
\item \textbf{RQ4}: How does MMGRec compare with traditional paradigm methods regarding inference efficiency?
\item \textbf{RQ5}: How do Rec-ID design choices affect collisions and recommendation performance?
\end{itemize}

\subsection{Performance Comparison (RQ1)}
Table~\ref{tab:performance} summarizes the performance of MMGRec and the baseline methods on three experimental datasets. It also reports the relative improvements and statistical significance tests, computed between our model and the strongest baseline (underlined). From these results, we make the following observations:

\begin{itemize}[leftmargin=*]
\item MMGRec consistently delivers the best performance across all settings. In particular, it outperforms the strongest competitor in terms of NDCG@10 by 7.17\%, 6.79\%, and 6.58\% on MovieLens, TikTok, and Kwai, respectively. We further perform one-sample t-tests, which indicate that the gains achieved by MMGRec are statistically significant (p-value $<$ 0.05). The key distinction between MMGRec and the baseline methods lies in the adoption of a generative rather than a conventional recommendation paradigm. Therefore, we attribute the substantial performance improvements to the introduction of a Transformer-based generative framework for multimodal recommendation.
\item MMGRec and the strongest baseline LightGT outperform other baselines by a large margin. This phenomenon validates that the Transformer is applicable to the multimodal recommendation. Nevertheless, LightGT focuses on improving representation learning by only utilizing the Transformer encoder. In comparison, our model exploits the generative capability of the complete Transformer, thus achieving superior performance.
\item Comparing the performance of MMGRec on two evaluation metrics, we find larger improvements in terms of NDCG. Namely, MMGRec's recommendation result is more accurate with respect to preference ranking. This illustrates that the generative paradigm is especially good at inferring the user's favorite items.
\end{itemize}

\begin{table}[!tbp]\centering
\caption{Overall performance comparison between our model and the baselines on three datasets.}
\setlength{\tabcolsep}{1.0mm}{
\begin{tabular}{l|cc|cc|cc}
\hline
\multirow{2}*{Methods} & \multicolumn{2}{c|}{MovieLens} & \multicolumn{2}{c|}{TikTok} & \multicolumn{2}{c}{Kwai}\\
\cline{2-7}
{} & Recall@10 & NDCG@10 & Recall@10 & NDCG@10 & Recall@10 & NDCG@10 \\
\hline
\hline
GraphSAGE & 0.2129 & 0.1388 & 0.0778 & 0.0476 & 0.0424 & 0.0344\\
NGCF & 0.2340 & 0.1383 & 0.0906 & 0.0547 & 0.0427 & 0.0363\\
GAT & 0.2342 & 0.1589 & 0.0945 & 0.0575 & 0.0441 & 0.0369 \\
LightGCN & 0.2381 & 0.1592 & 0.0988 & 0.0603 & 0.0502 & 0.0411\\
\hline
VBPR & 0.1927 & 0.1207 & 0.0600 & 0.0397 & 0.0302 & 0.0221 \\
MMGCN & 0.2453 & 0.1523 & 0.0645 & 0.0579 & 0.0456 & 0.0374 \\
GRCN & 0.2520 & 0.1683 & 0.0952 & 0.0584 & 0.0473 & 0.0403 \\
LATTICE & 0.2520 & 0.1680 & 0.0984 & 0.0611 & 0.0483 & 0.0400 \\
InvRL & 0.2518 & 0.1666 & 0.1002 & 0.0612 & 0.0486 & 0.0402\\
LightGT & \underline{0.2650} & \underline{0.1771} & \underline{0.1213} & \underline{0.0751} & \underline{0.0546} & \underline{0.0441}\\
\hline
\textbf{MMGRec} & \textbf{0.2804} & \textbf{0.1898} & \textbf{0.1269} & \textbf{0.0802} & \textbf{0.0567} & \textbf{0.0470}\\
\hline
\hline
\% Improv. & 5.81\% & 7.17\% & 4.62\% & 6.79\% & 3.85\% & 6.58\%\\
p-value & 8.15e-4 & 2.36e-4 & 3.83e-2 & 3.14e-3 & 5.70e-3 & 1.27e-2\\
\hline
\end{tabular}
}
\label{tab:performance}
\end{table}

\subsection{Ablation Study (RQ2)}

\begin{table}[!tbp]\centering
\caption{Performance comparison between our model and the variants without item representations.}
\setlength{\tabcolsep}{0.8mm}{
\begin{tabular}{l|cc|cc|cc}
\hline
\multirow{2}*{Methods} & \multicolumn{2}{c|}{MovieLens} & \multicolumn{2}{c|}{TikTok} & \multicolumn{2}{c}{Kwai}\\
\cline{2-7}
{} & Recall@10 & NDCG@10 & Recall@10 & NDCG@10 & Recall@10 & NDCG@10 \\
\hline
\hline
ID w/o & 0.1934 & 0.1238 & 0.0811 & 0.0516 & 0.0417 & 0.0361\\
Emb w/o & 0.2780 & 0.1890 & 0.1206 & 0.0792 & 0.0553 & 0.0460\\
Ours & \textbf{0.2804} & \textbf{0.1898} & \textbf{0.1269} & \textbf{0.0802} & \textbf{0.0567} & \textbf{0.0470}\\
\hline
\end{tabular}}
\label{tab:init}
\end{table}

\subsubsection{\textbf{Effect of Item Representation}}
To verify the necessity of item representations for MMGRec, we adopt two variants:

\textbf{ID w/o}: Without (w/o) item representation, the variant concatenates multimodal features as the input of RQ-VAE to assign Rec-ID.

\textbf{Emb w/o}: The variant replaces item representations in the Transformer encoder's input embedding with parameter embeddings.

Table~\ref{tab:init} displays the performance comparison between MMGRec and its variants, and we have the following observations:
\begin{itemize}[leftmargin=*]
\item The poor performance of \textbf{ID w/o} suggests that heterogeneous multimodal features cannot be directly used for Rec-ID assignment. The likely reason is the complexity of semantic relations among different modalities, necessitating prior integration through multimodal fusion. This aligns with the prior study about the importance of multimodal fusion in semantic comprehension~\cite{baltruvsaitis2018multimodal}.
\item Despite parameterizing input embedding, \textbf{Emb w/o} underperforms and is inferior to utilizing well-trained item representations.  Hence, we infer that the Transformer struggles to learn suitable input embeddings directly from the generative recommendation task without pre-training. This is consistent with many Transformer-based NLP models that benefit from pre-trained word embedding like Word2Vec and Glove.
\end{itemize}

\begin{table}[!tbp]\centering
\caption{Performance comparison among different Rec-ID assignment approaches.}
\setlength{\tabcolsep}{0.8mm}{
\begin{tabular}{l|cc|cc|cc}
\hline
\multirow{2}*{Methods} & \multicolumn{2}{c|}{MovieLens} & \multicolumn{2}{c|}{TikTok} & \multicolumn{2}{c}{Kwai}\\
\cline{2-7}
{} & Recall@10 & NDCG@10 & Recall@10 & NDCG@10 & Recall@10 & NDCG@10 \\
\hline
\hline
HK-Means & 0.2520 & 0.1683 & 0.1137 & 0.0742 & 0.0469 & 0.0402\\
Random & 0.2762 & 0.1872 & 0.1199 & 0.0785 & 0.0544 & 0.0452\\
Ours & \textbf{0.2804} & \textbf{0.1898} & \textbf{0.1269} & \textbf{0.0802} & \textbf{0.0567} & \textbf{0.0470}\\
\hline
\end{tabular}}
\label{tab:sid}
\end{table}

\subsubsection{\textbf{Effect of Graph RQ-VAE}}
To study the importance of Graph RQ-VAE, we compare it with Hierarchical K-Means Clustering (\textbf{HK-Means})~\cite{chen2005novel} in Rec-ID assignment. HK-Means aims at establishing a cluster hierarchy, employing a top-down methodology that initiates with a single cluster including all items, then iteratively divides it into smaller clusters according to item representations. The hierarchical cluster indices of an item can be concatenated as the Rec-ID.
Experimentally, we adjust the clustering number to obtain optimal performance and comparable cardinality.

\begin{itemize}[leftmargin=*]
\item Table~\ref{tab:sid} shows the performance comparison between employing Graph RQ-VAE and HK-Means. Graph RQ-VAE consistently outperforms HK-Means, illustrating the advantageous quantization ability. To explore the reason, we find that more ID collisions exist in HK-Means due to the clustering characteristic.
This illustrates that collisions negatively affect performance and should be handled reasonably. 
\item Therefore, we also compare our collision solution with the previous method (denoted by \textbf{Random}) that randomly extends one token to make Rec-ID unique. As the results recorded in Table~\ref{tab:sid}, it is obvious that such a random method is consistently inferior to our solution. This verifies that the Transformer cannot accurately generate the last random token, while our method relatively circumvents this weakness.
\end{itemize}

\subsubsection{\textbf{Effect of Relation-Aware Self-Attention}}
To investigate whether MMGRec can benefit from relation-aware self-attention, we consider two variants of our model:

\textbf{Default PE}: This variant preserves the default sinusoid-based positional encoding of Transformer. 

\textbf{w/o PE}: This variant entirely discards positional encoding. 

Note that our investigation of positional encoding is limited to the Transformer's encoder, while the decoder's requires no changes. Table~\ref{tab:pos} exhibits MMGRec's performance with different position encoding approaches. We find that:
\begin{itemize}[leftmargin=*]
\item The variant without any positional encoding achieves poor performance. Even using absolute positions of input elements for positional encoding can improve the performance. This indicates the importance of positional information for the Transformer~\cite{shaw2018self}. 
\item Compared to default positional encoding, relation-aware self-attention yields remarkable improvements, suggesting the effectiveness of leveraging pairwise item relations as positional information. 
This also validates the necessity of utilizing user information to model user-specific pairwise item relations, distinguishing our design from a standard self-attention head.
\end{itemize}

\begin{table}[!tbp]\centering
\caption{Performance comparison among different positional encoding approaches.}
\setlength{\tabcolsep}{0.8mm}{
\begin{tabular}{l|cc|cc|cc}
\hline
\multirow{2}*{Methods} & \multicolumn{2}{c|}{MovieLens} & \multicolumn{2}{c|}{TikTok} & \multicolumn{2}{c}{Kwai}\\
\cline{2-7}
{} & Recall@10 & NDCG@10 & Recall@10 & NDCG@10 & Recall@10 & NDCG@10 \\
\hline
\hline
w/o PE & 0.2696 & 0.1850 & 0.1152 & 0.0765 & 0.0517 & 0.0427\\
Default PE & 0.2732 & 0.1853 & 0.1180 & 0.0778 & 0.0545 & 0.0441\\
Ours & \textbf{0.2804} & \textbf{0.1898} & \textbf{0.1269} & \textbf{0.0802} & \textbf{0.0567} & \textbf{0.0470}\\
\hline
\end{tabular}
}
\label{tab:pos}
\end{table}

\begin{figure}[tbp]
\centering
\begin{minipage}[t]{0.4\linewidth}
\centering
\includegraphics[width=\linewidth]{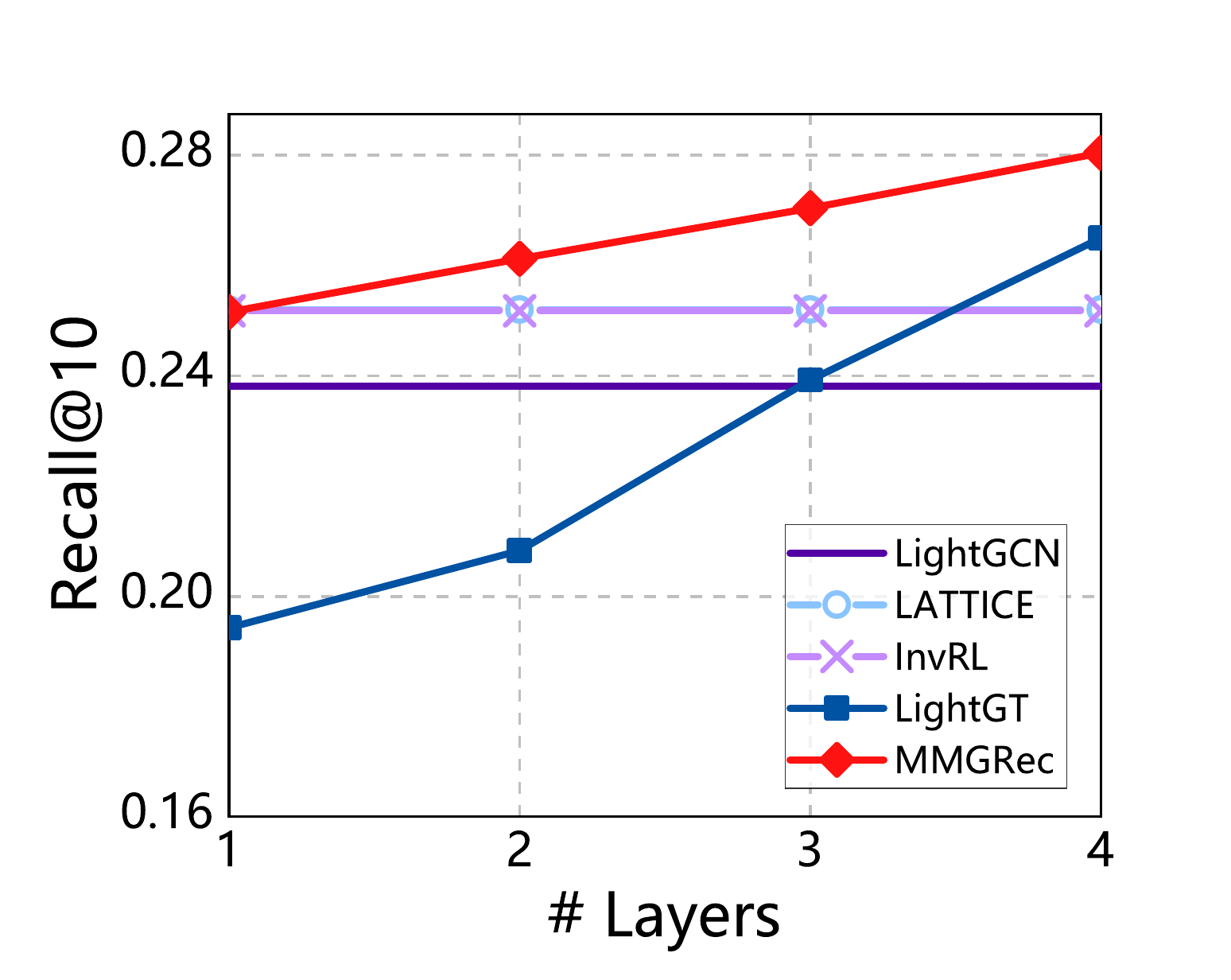}
\end{minipage}
\begin{minipage}[t]{0.4\linewidth}
\centering
\includegraphics[width=\linewidth]{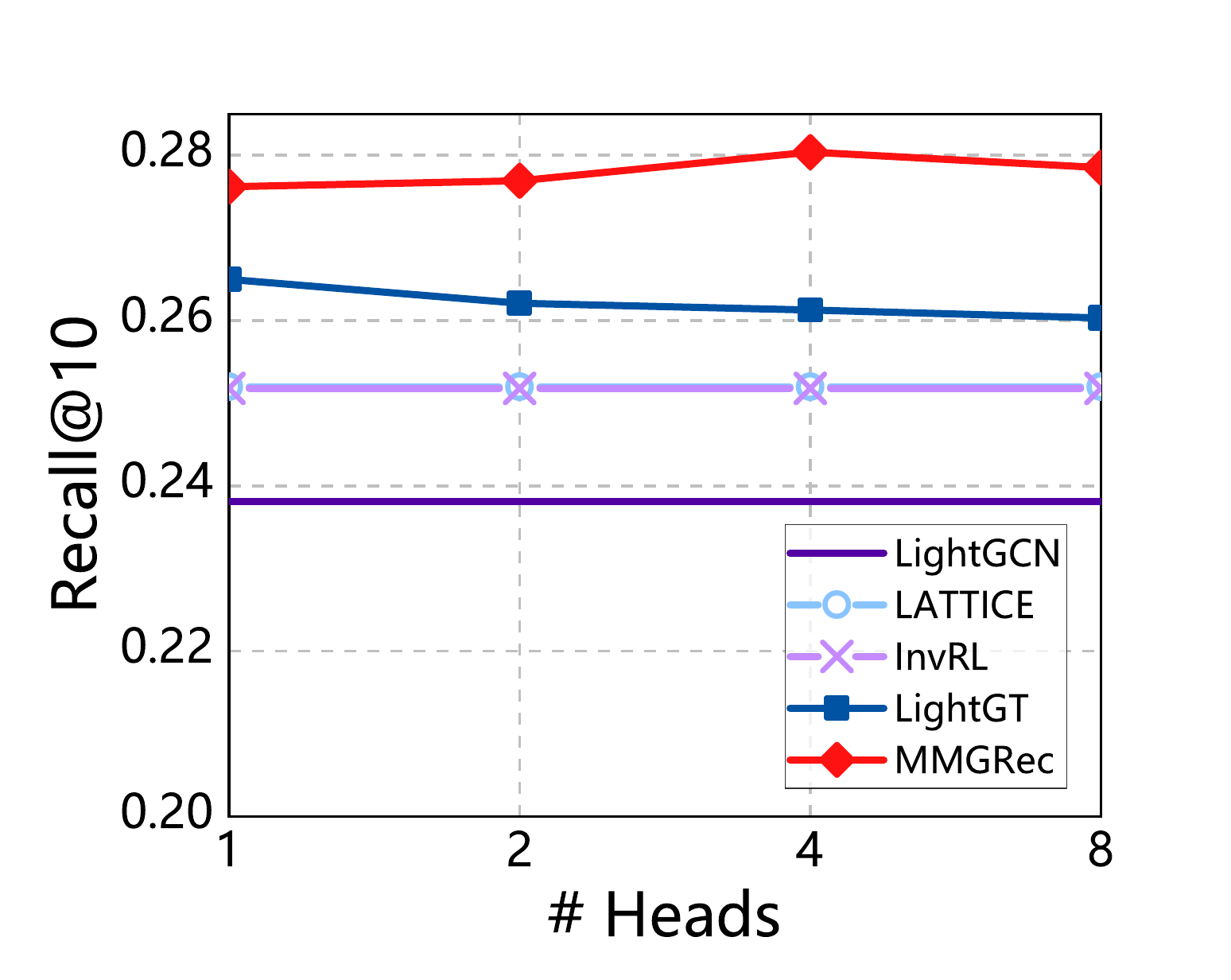}
\end{minipage}
\centerline{(a)~MovieLens}
\\
\begin{minipage}[t]{0.4\linewidth}
\centering
\includegraphics[width=\linewidth]{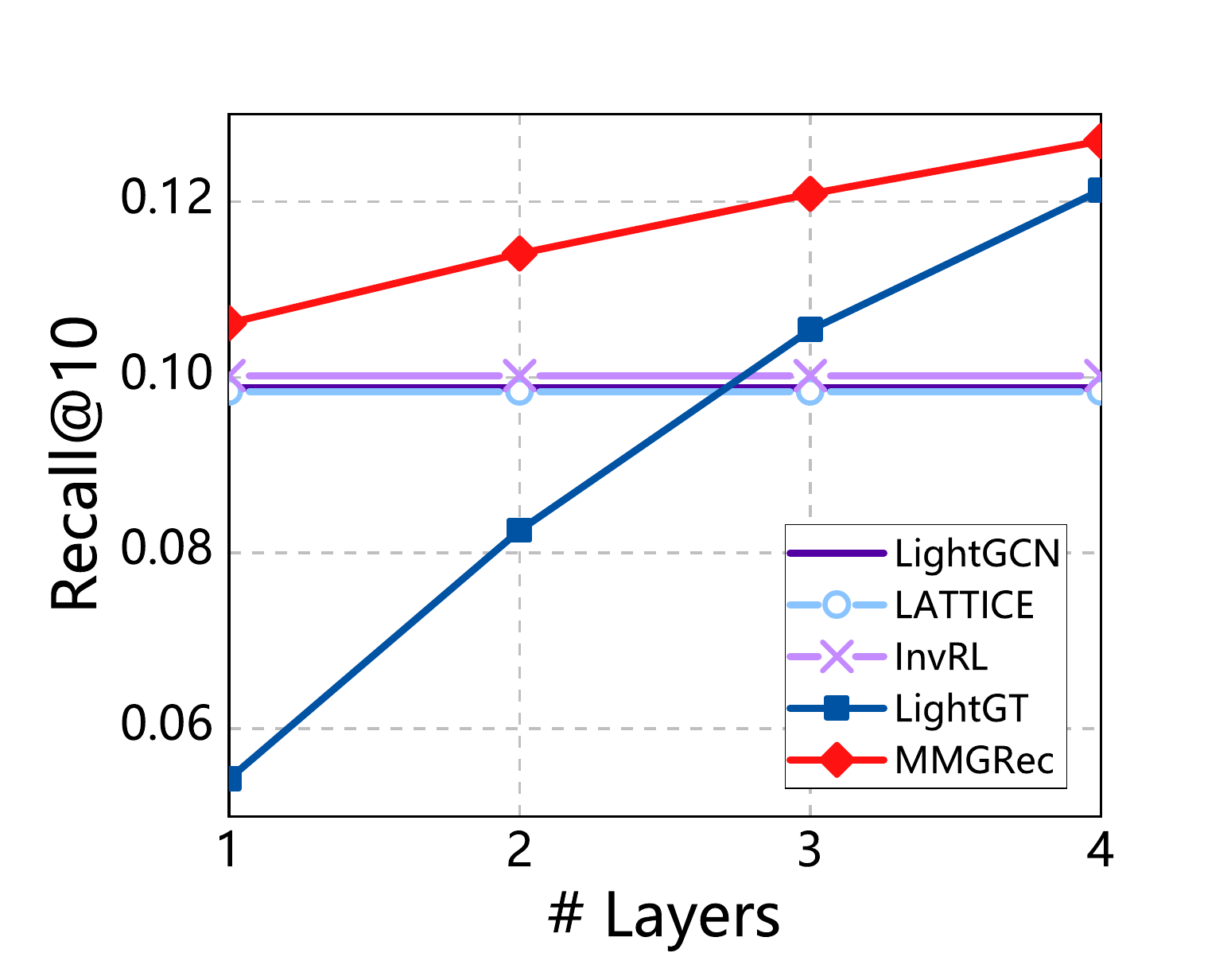}
\end{minipage}
\begin{minipage}[t]{0.4\linewidth}
\centering
\includegraphics[width=\linewidth]{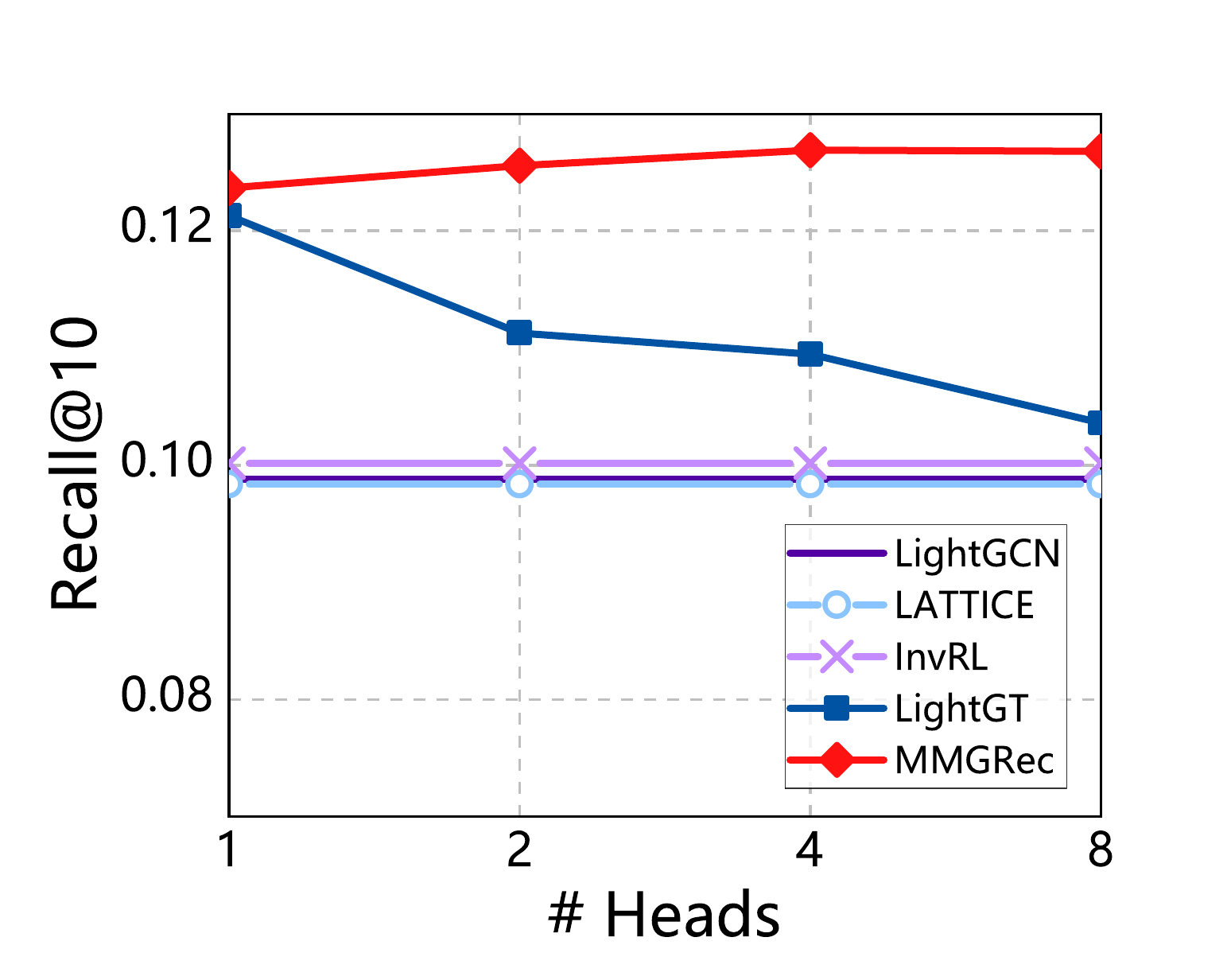}
\end{minipage}
\centerline{(b)~TikTok}
\\
\begin{minipage}[t]{0.4\linewidth}
\centering
\includegraphics[width=\linewidth]{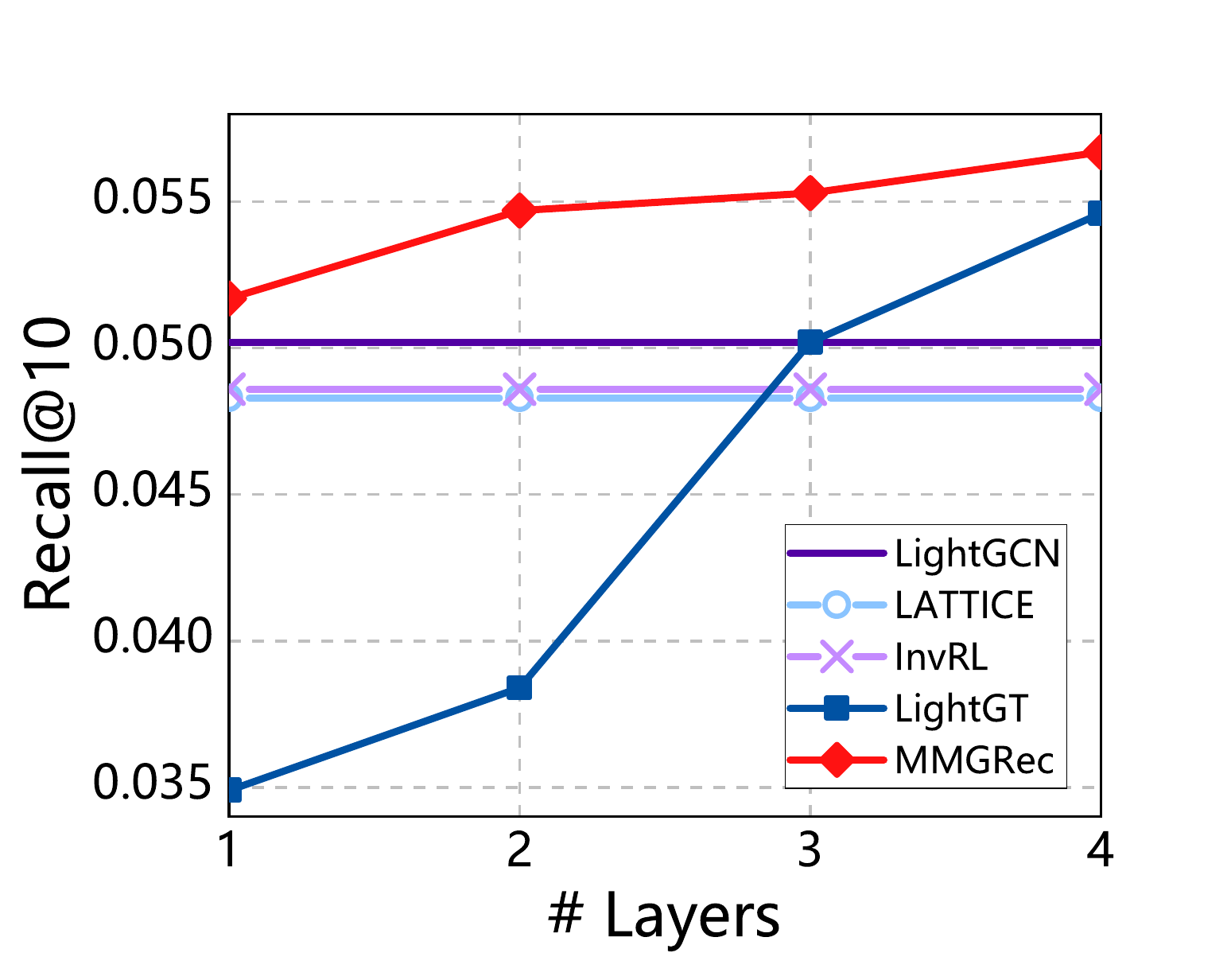}
\end{minipage}
\begin{minipage}[t]{0.4\linewidth}
\centering
\includegraphics[width=\linewidth]{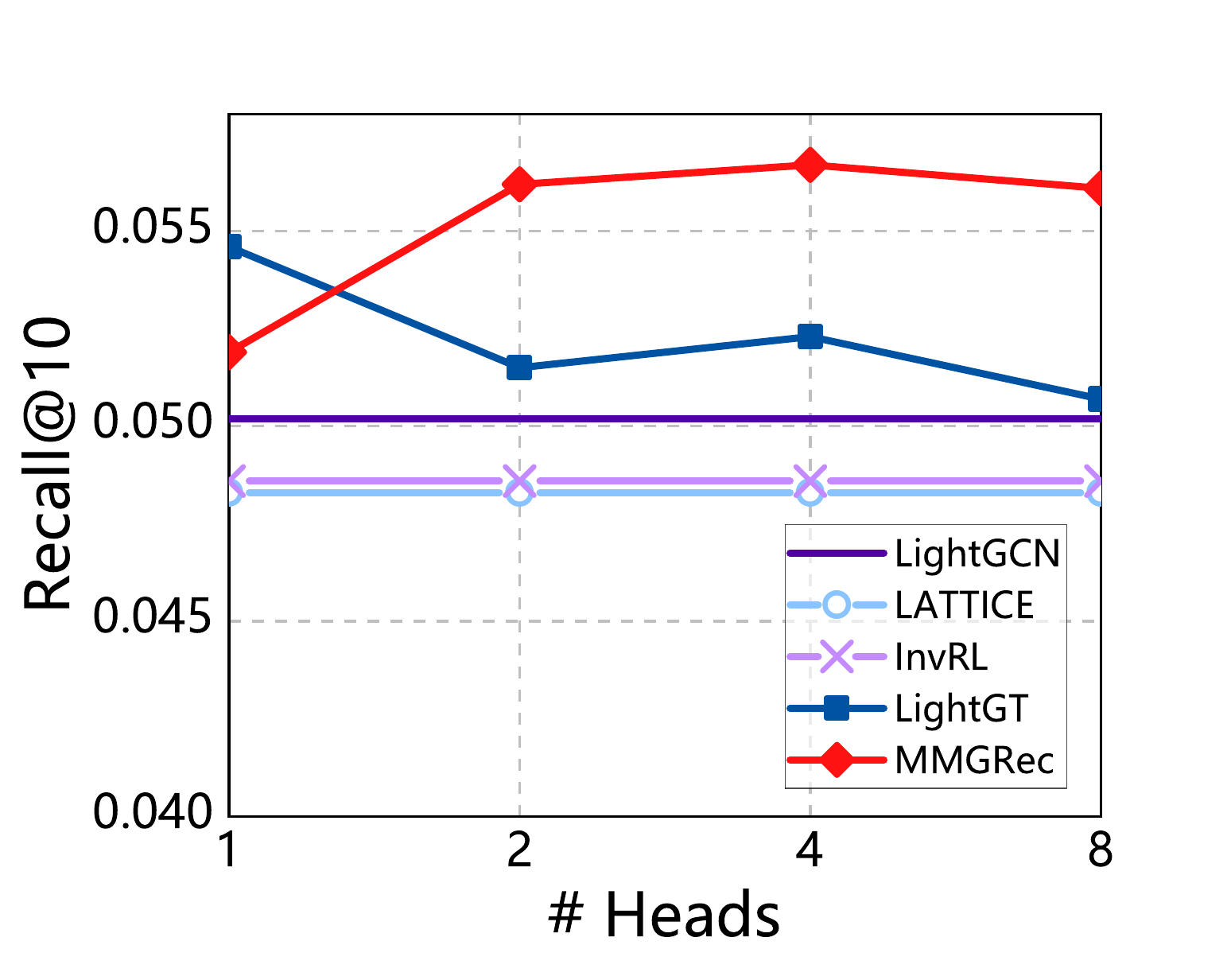}
\end{minipage}
\centerline{(c)~Kwai}
\centering
\caption{Effect of layer and head numbers.}
\label{fig:layer_head}
\end{figure}

\subsection{Hyper-Parameter Analysis (RQ3)}

To investigate the impact of Transformer layers on our model, we vary the layer number in the range of \{1, 2, 3, 4\} when fixing the self-attention head number to optimal. As a comparison, we test the LightGT with the same setting. Left of Figure~\ref{fig:layer_head} plots the results \textit{w.r.t.} Recall@10 on three datasets, we observe that: First, MMGRec outperforms LightGT as varying layer numbers across all datasets. This validates the stable superiority derived from the generative recommendation. Second, MMGRec can achieve quite good performance with fewer layers. This indicates that our generative model can be effective in a fundamental setting, while the embedding-based method needs stacking Transformer layers to guarantee performance.

Moreover, we investigate the impact of multi-head attention on our model by comparing 1-head, 2-head, 4-head, and 8-head self-attention blocks with LightGT. As shown in the right of Figure~\ref{fig:layer_head}, we observe that MMGRec consistently outperforms LightGT. As the head number increases, the performance of MMGRec tends to stabilize after rising. In comparison, the performance of LightGT shows a continued downward trend. Theoretically, one head of self-attention models the relation among elements in a single aspect. Therefore, we infer that the generative model can handle more kinds of relations than the embedding-based method.


\subsection{Efficiency Study (RQ4)}
\begin{figure}[t]
    \centering
    \includegraphics[width=0.3\linewidth]{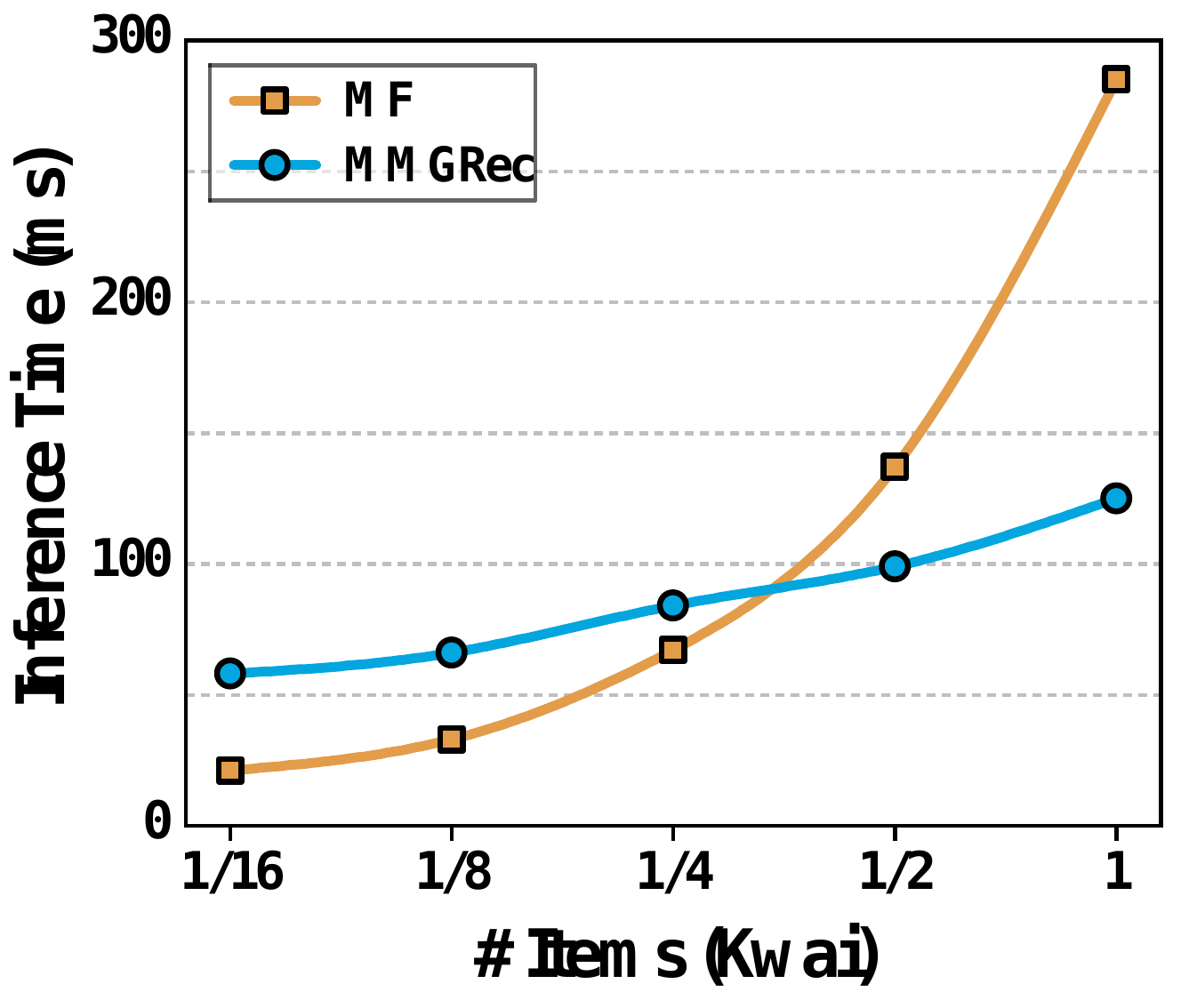}
    \includegraphics[width=0.3\linewidth]{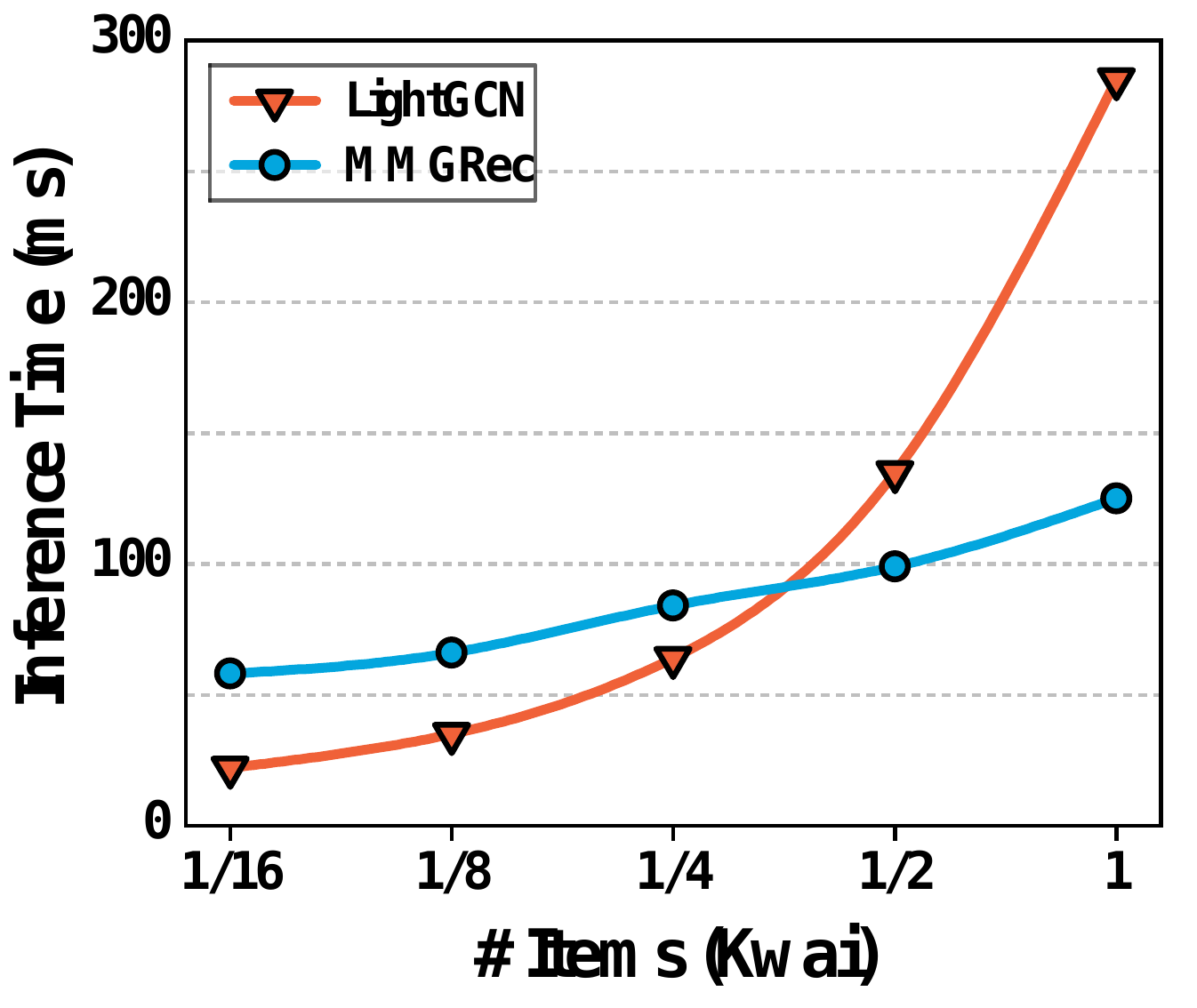}
    \includegraphics[width=0.3\linewidth]{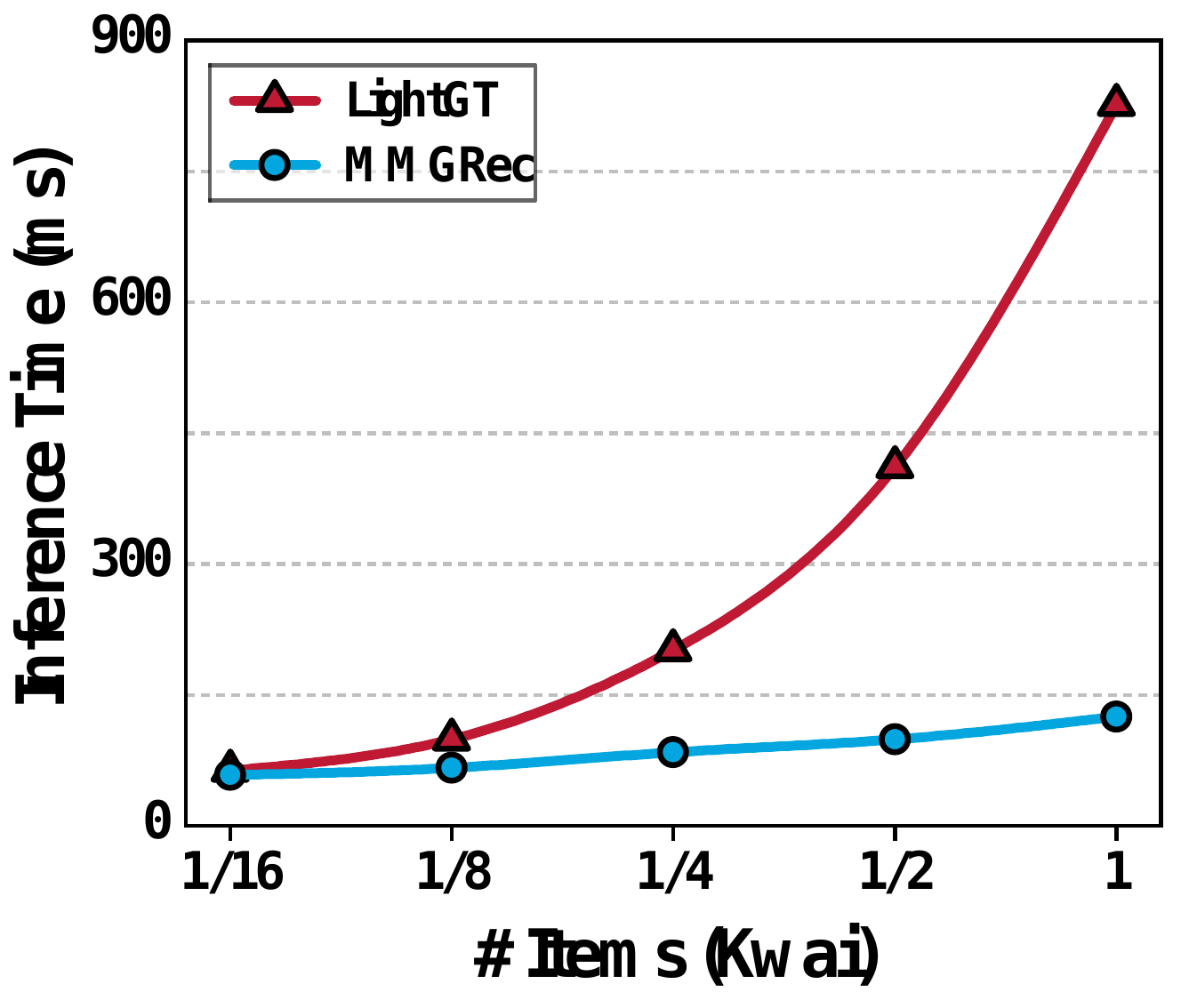}
    \caption{Inference time on different-scale Kwai dataset. The unit of time is milliseconds (ms).}
    \label{fig:infertime}
\end{figure}

Instead of computing and ranking similarity, MMGRec employs autoregressive decoding to directly generate Rec-IDs identifying items during inference. To assess the efficiency of the two inference approaches, we conduct comparative experiments using identical computing configurations and record their inference time.
Figure~\ref{fig:infertime} presents a pairwise efficiency comparison between MMGRec and each of MF, LightGCN, and LightGT, in terms of the average inference time per user across Kwai datasets of varying scales.
\textbf{1/8 Kwai} denotes randomly selecting one in eight items from the complete dataset. The main observations are as follows:

For LightGCN and LightGT, the propagated embeddings are precomputed offline, such that the online inference is effectively reduced to inner-product retrieval. Under this setting, the computational complexity of LightGCN is comparable to that of MF, while LightGT exhibits higher complexity due to inner-product computations across multiple modalities. Consequently, our efficiency analysis primarily focuses on comparing MMGRec with MF, which represents the most efficient baseline.
In scenarios where the item count is low (\textit{e.g.}, $1/16\sim 1/4$ Kwai), MMGRec exhibits inefficiency, primarily due to the basic computational demands of autoregressive decoding. However, as the number of items increases (\textit{e.g.}, $1/4\sim 1$ Kwai), the inference time of MF experiences a significant rise. This escalation is attributed to the fact that item count linearly impacts the time complexity of inner product computation and sorting. In contrast, MMGRec demonstrates enhanced efficiency as its inference time remains relatively stable. This stability arises from the unchanged number of autoregressive decoding steps, even with an increase in the number of items. These findings suggest that the generative paradigm holds the potential to deliver more efficient inference for large-scale recommendation.

\subsection{Analysis of Rec-ID Design (RQ5)}
\begin{figure}[t]
    \centering
    \begin{minipage}{0.329\linewidth}
        \centering
        \includegraphics[width=\linewidth]{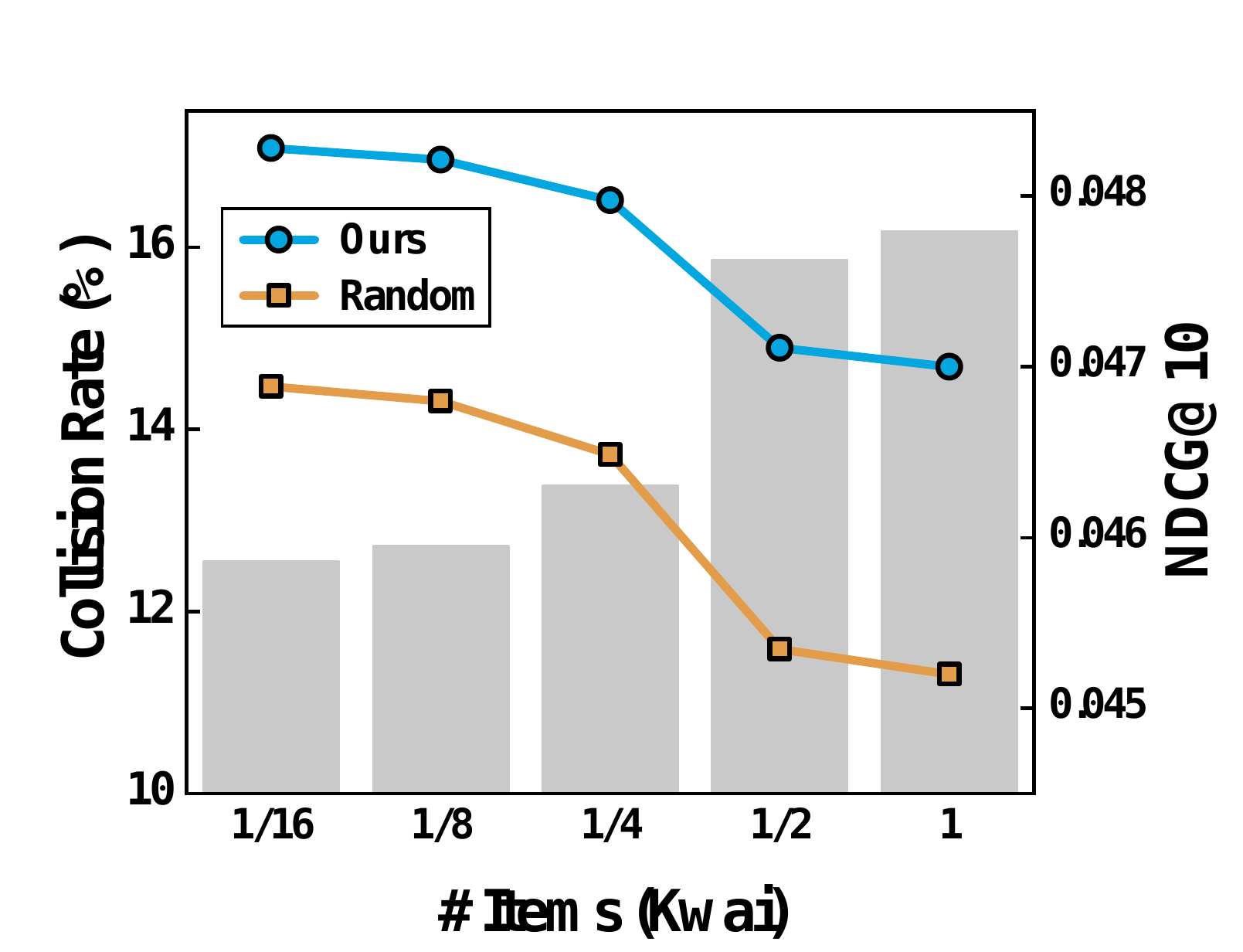}\\[-1mm]
        (a)
    \end{minipage}
    \begin{minipage}{0.329\linewidth}
        \centering
        \includegraphics[width=\linewidth]{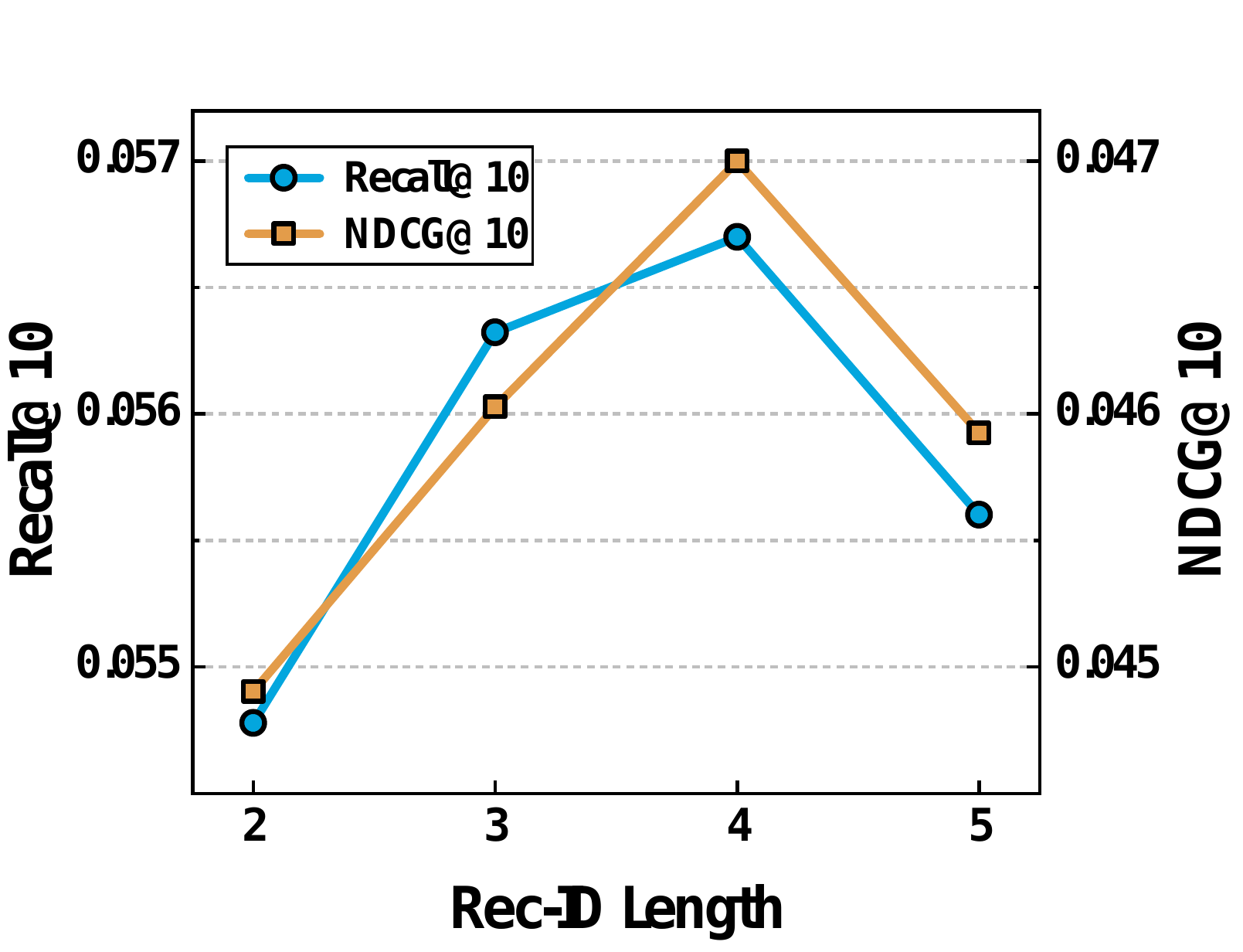}\\[-1mm]
        (b)
    \end{minipage}
    \begin{minipage}{0.329\linewidth}
        \centering
        \includegraphics[width=\linewidth]{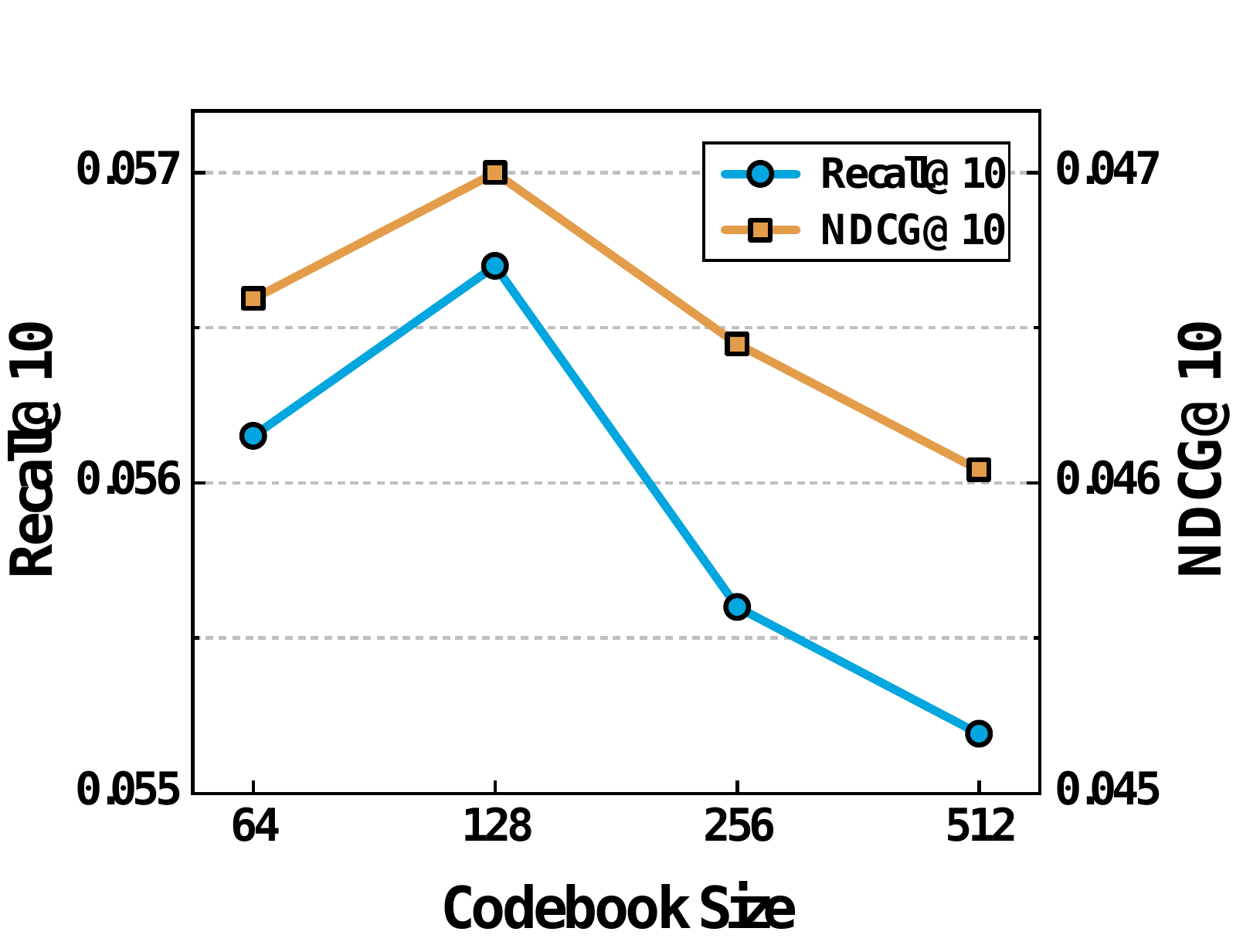}\\[-1mm]
        (c)
    \end{minipage}
    \caption{Quantitative analysis of Rec-ID design on the Kwai dataset. (a) Impact of Rec-ID collision and popularity token. (b) Effect of Rec-ID length. (c) Effect of codebook size.}
    \label{fig:exp-rec-id}
\end{figure}

This subsection provides a quantitative analysis of the Rec-ID design, focusing on the effect of the popularity token in mitigating Rec-ID collisions, as well as the impact of Rec-ID length and codebook size on model performance. Due to consistent trends across datasets and space limitations, we report results only on the Kwai dataset, as shown in Figure~\ref{fig:exp-rec-id}.

\textit{\textbf{Impact of Rec-ID Collision.}}
As shown in Figure~\ref{fig:exp-rec-id}(a), the bar plots show the collision rate on the Kwai dataset under different data scales, measured as the percentage of items assigned non-unique Rec-IDs. The x-axis value \textbf{1/16} indicates a subset sampled from one-sixteenth of the items and their interacting users. As the dataset scale increases, the collision rate rises, and model performance degrades slightly, indicating that Rec-ID collisions have a non-negligible impact in large-scale recommendation scenarios. The line plots compare our popularity token with a random token (as in TIGER). With increasing collision rates, the popularity token outperforms the random token by a larger margin. This indicates that popularity-aware semantic ordering better distinguishes colliding Rec-IDs and reduces the performance degradation caused by collisions.

\textit{\textbf{Effect of Rec-ID Length.}}
Figure~\ref{fig:exp-rec-id}(b) shows the performance trends in Recall@10 and NDCG@10 as the Rec-ID length $M$ increases from 2 to 5. To cover the full item set, a shorter Rec-ID results in a larger per-step decoding search space. As $M$ increases from 2 to 4, model performance steadily improves, since overly short Rec-IDs increase decoding difficulty due to a large set of candidate tokens at each step. When $M=4$, the model achieves a good balance between the number of decoding steps and the per-step search space, leading to the best performance. However, when $M$ further increases to 5, more autoregressive steps introduce higher generation difficulty and inference latency, resulting in performance degradation~\cite{wang2024content}. Therefore, we set $M=4$ by default.

\textit{\textbf{Effect of Codebook Size.}}
To evaluate the impact of codebook size $L$, we fix the Rec-ID length to $M=4$ and vary $L$ over $\{64,128,256,512\}$. As shown in Figure~\ref{fig:exp-rec-id}(c), performance improves as $L$ increases from 64 to 128, since a small codebook has limited representational capacity and cannot adequately distinguish Rec-ID semantics. When $L=128$, the codebook is sufficiently expressive and achieves the best performance. However, further increasing $L$ rapidly enlarges the decoding search space, forcing the model to select from more candidate tokens at each step, which increases generation uncertainty and degrades performance. These results suggest that codebook size should balance representational capacity and decoding complexity in generative recommendation.

\section{Conclusion}
In this work, we propose MMGRec, a novel Transformer-based model for multimodal recommendation. MMGRec explores the generative paradigm to address limitations inherent in the traditional paradigm, consisting of two essential components --- Rec-ID assignment and Rec-ID generation. In Rec-ID assignment, we integrate item multimodal data with CF information and develop a Graph RQ-VAE to quantize them into Rec-ID for each item. In Rec-ID generation, a Transformer model is trained to directly predict item Rec-IDs for recommendation based on a user's interaction sequence. Experimental results demonstrate that MMGRec achieves state-of-the-art performance while maintaining conditionally efficient inference.

For future work, we plan to enhance MMGRec's performance by developing a more effective vector quantization method to mitigate Rec-ID assignment collisions. Additionally, we intend to explore the integration of large models~\cite{hua2023tutorial}, such as ChatGPT, to improve user and item representations. Furthermore, we are interested in leveraging MMGRec's generative capability to tackle the cold-start problem in recommendation~\cite{sun2020lara}.

\begin{acks}
This work was supported by the National Natural Science Foundation of China (NSFC; No. 62572282); the Guangdong S\&T Programme (No. 2025B0101130003); the Key R\&D Program of Shandong Province, China (No. 2025CXGC020101); NSFC (Nos. U24A20328 and 62476071); the Guangdong Basic and Applied Basic Research Foundation (No. 2025A1515011732); and NSFC (No. 62376137).
\end{acks}

\bibliographystyle{ACM-Reference-Format}
\bibliography{samples/reference}


\end{document}